\documentclass[aoas]{imsart}

\RequirePackage{amsthm,amsmath,amsfonts,amssymb}
\RequirePackage[authoryear]{natbib}
\RequirePackage[colorlinks,citecolor=blue,urlcolor=blue]{hyperref}
\RequirePackage{graphicx}

\startlocaldefs
\newcommand{\ind}{\perp\!\!\!\!\perp}
\endlocaldefs

\begin{document}

\begin{frontmatter}
\title{Evaluating the effects of policy interventions subject to early adoption: A case study of prescription drug monitoring programs and opioid dispensing}
\runtitle{Policy interventions subject to early adoption}

\begin{aug}
\author[A]{\fnms{Sarika}~\snm{Aggarwal}\ead[label=e1]{saggarwal@g.harvard.edu}}
\author[A]{\fnms{Brent A.}~\snm{Coull}\ead[label=e2]{bcoull@hsph.harvard.edu}}
\author[A]{\fnms{Nima}~\snm{Hejazi}\ead[label=e3]{nhejazi@hsph.harvard.edu}}
\author[A]{\fnms{Rachel C.}~\snm{Nethery}\ead[label=e4]{rnethery@hsph.harvard.edu}}
\address[A]{Department of Biostatistics, Harvard T.H. Chan School of Public Health\printead[presep={,\ }]{e1}\printead[presep={,\ }]{e2}\printead[presep={,\ }]{e3}\printead[presep={,\ }]{e4}}
\end{aug}

\begin{abstract}
Policies that require organizations to use new systems, such as prescription drug monitoring programs (PDMPs), are often implemented in phases, with an initial period of voluntary access followed by mandated compliance. This allows the policy intervention to be adopted before compliance is required (early adoption), causing outcomes to change before the mandate takes effect. When early adoption is present, the no-anticipation assumption underlying synthetic control methods (SCM) is violated, leading to biased policy effect estimates. We formalize early adoption in a potential outcomes framework for staggered policy implementation and decompose the total policy effect into early adoption and mandate components. We then propose a two-stage, early adoption-aware SCM procedure that first estimates early adoption effects using an interactive fixed effects model fit to pre-mandate data and then residualizes outcomes before applying SCM variants to estimate mandate and total policy effects. Simulations, including settings with correlation between early adoption and latent factors, show reduced bias and improved uncertainty quantification relative to conventional SCM estimators. We apply the framework to state-level PDMP policies and per-capita opioid dispensing. After accounting for early adoption, estimates suggest reductions in opioid dispensing following PDMP availability and mandates; however, the estimates are imprecise and not statistically significant.
\end{abstract}

\end{frontmatter}

\section{Introduction}
The opioid epidemic in the United States has evolved through discrete waves of prescription and illicit opioid use, with the first wave driven largely by rapidly increased dispensing of high-potency prescription opioids such as hydrocodone and oxycodone. Deaths attributed specifically to prescription opioids rose from 3,442 in 1999 to 16,416 in 2020, a more than 4-fold increase \citep{guy2017, strickler2019, kaur2023}. In response to rising misuse, overdoses, and drug-related mortality, many states have implemented prescription drug monitoring programs (PDMPs) to improve and facilitate safer opioid prescribing practices. PDMPs are state-run databases that track controlled substance dispensing at the patient, prescriber, pharmacy, and/or drug levels. In doing so, authorized users can identify potentially unsafe opioid use and adjust prescribing accordingly (e.g., avoiding initiating or continuing opioids, tapering doses, limiting high-risk combinations such as concurrent benzodiazepines, or co-prescribing naloxone) \citep{us2015opioid, dsouza2024, puac2020}. In a systematic review of the effects of PDMP policies on prescription opioid-related outcomes in the United States, Puac-Polanco et al. found that many studies reported reductions in opioid prescribing and dispensing following the implementation of mandatory PDMP use policies while evidence regarding opioid-related overdoses and mortality was inconsistent \citep{puac2020}. 

However, prior studies of the effects of PDMP policies do not account for a nuance of their implementation. PDMP policies are typically implemented in two phases: an initial period in which the database becomes operational and accessible on a voluntary basis, followed months or years later by mandated prescriber use (i.e., legal requirements to check the database before prescribing controlled substances). Because the infrastructure is developed and available prior to mandated use, opioid dispensing may begin to decline not only after mandatory-use requirements take effect, but also during the earlier period when prescribers have voluntary access. During this period, prescribers can review PDMP records to assess patient histories and the safety and clinical need for opioid use. Over time, this decline is expected to slow and level off as remaining prescribing primarily reflects necessary opioid use \citep{kaur2023}.

\begin{figure}
    \centering
    \includegraphics[width=0.75\linewidth]{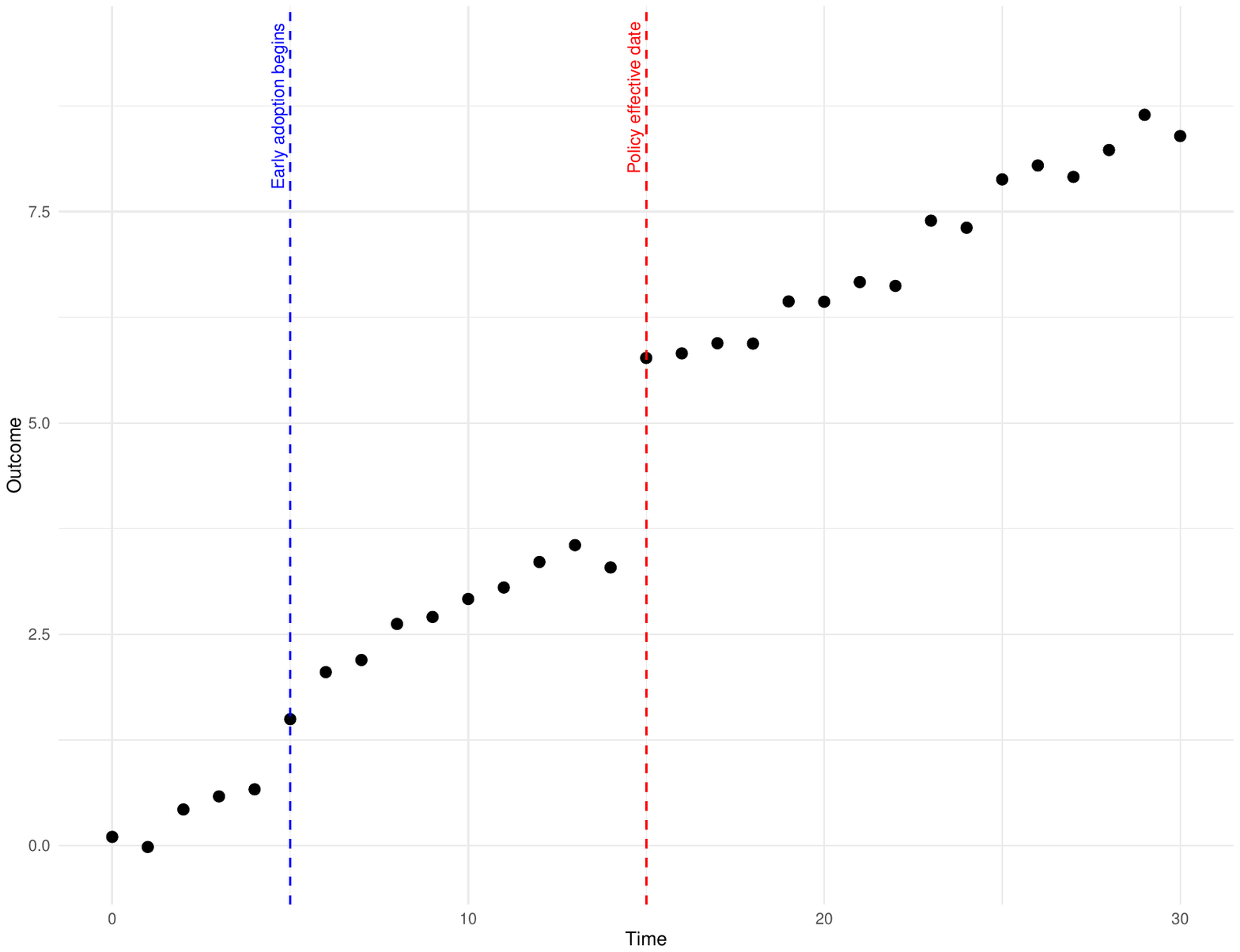}
    \caption{Visual example of early adoption. The blue vertical dashed line marks the onset of the early adoption period, indicated by a clear change in the outcome trajectory. The red vertical dashed line marks the onset of mandated compliance.}
    \label{visual_cue}
\end{figure}

We refer to such systematic pre-mandate changes in outcomes, which occur when infrastructure is established or made available before compliance is formally required, as \textit{early adoption} effects. Figure \ref{visual_cue} provides a toy illustration of early adoption effects in a panel data setting, showing a time series of outcomes that departs from its pre-policy trend before the policy's effective date as units begin to make voluntary use of new resources or infrastructure. The term ``anticipation effects'' has been used broadly in prior literature, typically referring to any policy-related forces that systematically impact the treated unit(s) prior to the policy's effective date. These may include the impacts of actions taken to prepare for the policy implementation/enforcement, actions taken to avoid anticipated regulation, or early adoption of policy-related infrastructure or practices \citep{augustin2025, piccininni2025}. In this paper, we focus on early adoption as one form of anticipation, guided by our motivating application, though our proposed approach to handle early adoption effects is flexible and may be appropriate for other types of anticipation effects as well. 

To evaluate the causal effects of public policies with panel data, quasi-experimental analytic methods are often leveraged. Among the most popular are the synthetic control method (SCM) family of methods, including the original SCM \citep{abadie2010} and several recent variants \citep{xu2017,ben2021}. In SCM, outcomes observed under control conditions (i.e., outcomes in control units and in treated units pre-treatment) are modeled as a function of observed covariates, unit and time fixed effects, and latent factors. Policy effects are estimated by comparing observed outcomes for treated units post-treatment with model-estimated counterfactual outcomes that would have been observed in the absence of treatment. A causal identifying assumption underlying the SCM framework requires that, conditional on observed covariates and the latent factor structure, the observed pre-treatment outcomes for treated units are consistent with the untreated potential outcomes up to the treatment time. Under this assumption, pre-treatment outcomes can be used to recover the untreated outcome path. This assumption prohibits anticipation effects, including early adoption effects. Under early adoption by some or all treated units, pre-treatment outcomes are affected by a version of treatment, and thus the untreated potential outcomes are not identified from the observed data. If naively applied ignoring the early adoption, SCM will fit to the observed pre-treatment path, which is ``contaminated'' and therefore recovers a counterfactual that does not correspond to the untreated potential outcomes. As a result, post-treatment effect estimates will generally be biased. Our simulations in Section~\ref{sec: simulation_study} show that, in the presence of early adoption, commonly used SCM estimators can substantially distort estimated policy effects. 

Little methodological research has addressed violations of the no-anticipation assumption. A commonly applied ad hoc approach is to discard observations from the pre-treatment period after anticipation effects are expected to have begun \citep{antwi2013}; however, this may not be practical when there is limited pre-treatment data or when the timing of behavioral changes is uncertain and potentially heterogeneous across groups. \citet{rambachan2023} highlight the lack of methodological guidance for addressing phenomena such as Ashenfelter's dip, where outcomes change prior to treatment due to selection bias or anticipation. More recently, researchers have proposed bounds on the average treatment effect on the treated (ATT) under different assumptions on the magnitude of anticipation effects \citep{billinski2024, gong2021}, which may be difficult to justify in practice and could result in bounds too wide to be useful for policy decision-making. Related work has shown that, in staggered adoption difference-in-differences designs with time-varying misclassification of treatment status and anticipation, the identifying assumptions underlying standard estimators fail (so that these estimators no longer recover common causal parameters) and propose modified estimators and specification tests that are robust to these forms of misspecification \citep{augustin2025}. \citet{piccininni2025} discuss how the standard no-anticipation assumption is ambiguous because it conflates policy implementation with the earlier decision or announcement of the policy, and they propose a framework that, instead, distinguishes these interventions and provides corresponding identification results. However, their setting focuses on anticipation driven by policy announcements, where the announcement affects outcomes before the policy is implemented but treatment itself begins only at implementation. In our context, anticipation arises from early adoption, in which units may voluntarily adopt the treatment before the policy mandate takes effect. This setting requires methods that distinguish the effects of voluntary early adoption effects from the additional effects of the policy mandate once it is implemented. Following the ideas of \citet{doudchenko2016}, Ben Michael et al. \citep{ben2021} discussed applying SCM to residualized outcomes, where outcomes are first regressed on auxiliary covariates and then residualized prior to synthetic control construction. Our approach builds on this idea by residualizing outcomes with respect to estimated early adoption effects rather than auxiliary covariates.  

In this paper, we formalize the concept of early adoption in panel data contexts using the potential outcomes framework and propose a method that explicitly disentangles early adoption from mandate effects. We allow outcomes for treated units to depend both on the time since policy-related resources become available and the time since the policy mandate takes effect as well as a latent factor structure that captures time-varying unobserved confounding. We then propose a two-stage estimation procedure where (1) the first stage disentangles the early adoption effects using pre-treatment data, and (2) the second stage subtracts out the estimated early adoption effects from the pre-treatment outcomes and uses these early adoption-free outcomes to estimate the treatment effects. 
This approach can handle settings with multiple treated units and staggered early adoption and treatment timing, while reducing bias arising from unmodeled early adoption.

Our contributions are summarized here. First, we provide a formal definition of early adoption and differentiate early adoption (pre-mandate), mandate, and total policy effects in staggered adoption panel data, showing how these quantities relate to commonly reported estimands. Second, we introduce a novel procedure to estimate each of the quantities above. Third, through simulation studies, we evaluate how various SCM estimators behave when early adoption is present but ignored, and compare to the performance of our proposed early adoption-aware estimator. We apply our proposed approach in our motivating application to assess the effects of PDMP policies on opioid dispensing patterns in the United States during the period 2000-2016. 

The remainder of the paper is organized as follows. Section 2 introduces the motivating policy data on state-level PDMP policies and opioid dispensing data. Section 3 describes the panel data structure, reviews the potential outcomes framework, and defines the three causal effects of interest: early adoption, mandate, and total effects along with the necessary identification assumptions. We also present our early adoption-adjusted estimation procedure and its implementation with two different SCM variants. Section 4 reports results from simulation studies under a range of data generating scenarios and Section 5 presents results from the real data analysis. Lastly, Section 6 concludes with a discussion and directions for future work.

\section{Motivating data}\label{sec:data}
\begin{figure}
    \centering
    \includegraphics[width=\linewidth]{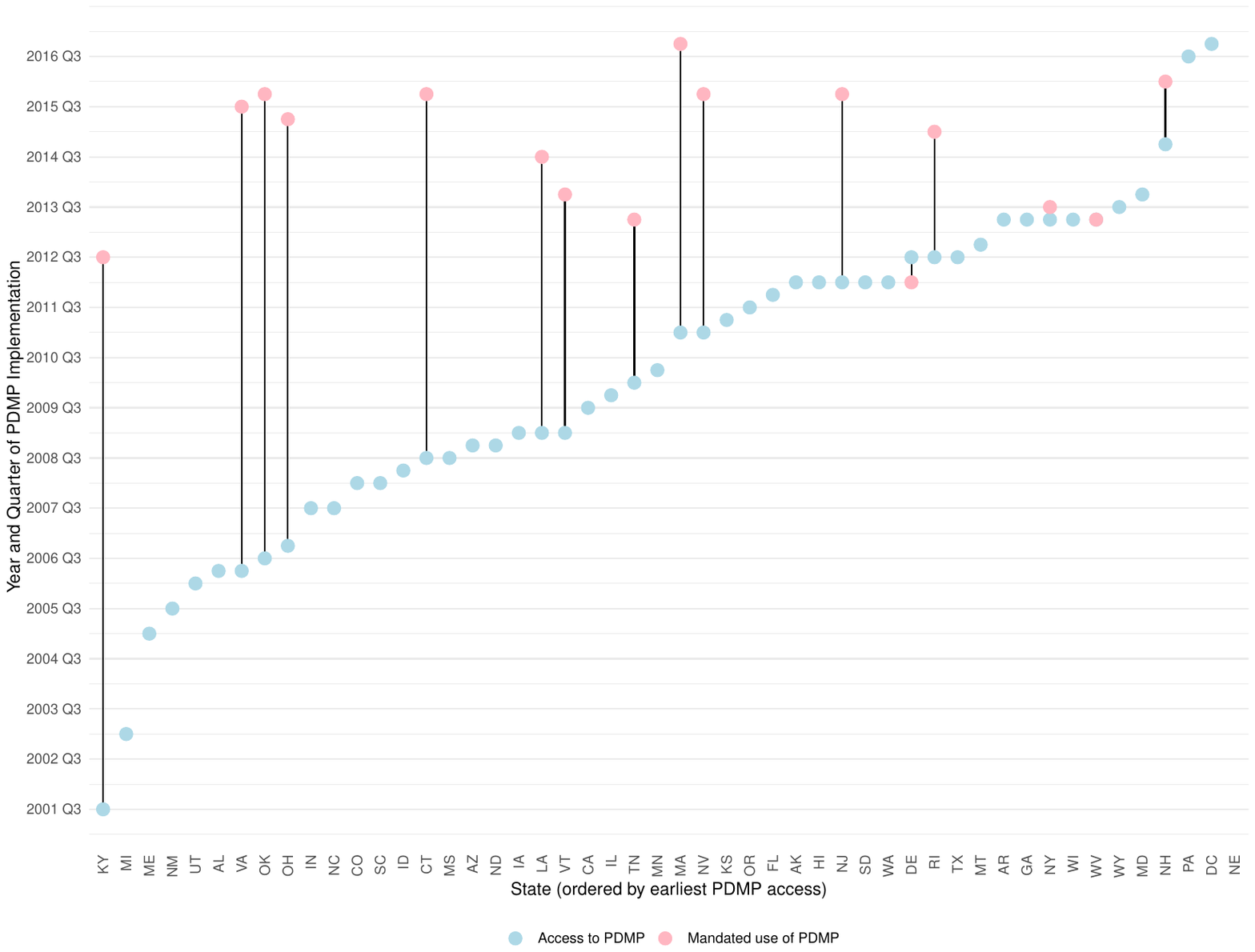}
    \caption{Timing of PDMP access and mandated use by state from 2000-2016. States are on the x-axis, ordered by the quarter in which prescribers first obtained access to the PDMP (blue dots). Pink dots indicate the quarter in which mandated use of the PDMP took effect, with vertical black segments showing the time elapsed between operational access and mandated use. States with only a blue dot did not enact a mandated-use requirement during the study period.}
    \label{policy_plot}
\end{figure}

We obtained data on state-level PDMP policies from the database constructed by Lee et al. \citep{lee2021}. PDMPs are state-run database systems that collect and disseminate data on controlled substance prescriptions, allowing authorized users such as prescribers and pharmacists to monitor prescribing and dispensing patterns at the patient, provider, pharmacy, and drug levels. These systems are intended to support safer prescribing decisions by identifying situations in which opioid use may be avoided or more closely managed. While state-specific policies vary in design, access, and substance coverage, a key distinction is whether prescribers are required to consult the PDMP prior to prescribing or whether use is voluntary. In many states, policies begin with optional use and later expand to mandate prescriber queries. Accordingly, this dataset records the month and year in which PDMPs became operational and accessible in each state, as well as the month and year in which prescriber use became mandatory under specified circumstances (i.e., required queries prior to prescribing opioids). The authors compiled information from six sources and reconciled discrepancies in dates across them. A plot showing the operational access date and, when applicable, mandated requirement date for each state through 2016 is provided in Figure \ref{policy_plot}.

We obtained quarterly state-level opioid dispensing quantities from the United States Drug Enforcement Administration's (DEA) Automation of Reports and Consolidated Orders System (ARCOS) database for the years 2000-2016 \citep{DEA_ARCOS}. ARCOS is a publicly available database that tracks the distribution of Schedule I-IV controlled substances from manufacturers and distributors to retail purchasers across the United States. We used ARCOS Report 3, which reports the quantity of drugs distributed (in grams) by drug type, state, and quarter, standardized to population size. Using these data, we constructed our outcome of interest: the quarterly per-capita quantity of opioids, specifically hydrocodone and oxycodone, distributed in each state. 
We converted the total grams of each drug into morphine milligram equivalents (MME) using standard oral conversion factors \citep{conversion_factors}, and then computed the total MME across drugs to obtain a single measure of opioid dispensing in MME for each state-quarter. 

We linked the policy data to our quarterly opioid dispensing measure (MME per capita) by assigning policy timing at the quarterly level. Policy dates were recorded at the month-year level and mapped to the corresponding calendar quarter. Specifically, the quarter containing the PDMP operational (access) date defines the beginning of the early adoption period while the quarter containing the PDMP mandatory usage date marks the point at which the policy becomes effective (i.e., the mandate takes effect).  

\section{Methods}

\subsection{Review of SCM}\label{sec:review_of_scm}
Synthetic control methods (SCM) are a class of methods for panel data that estimate causal effects by constructing a counterfactual outcome trajectory for treated units post-treatment using information from control units and pre-treatment outcomes in the treated \citep{abadie2010,xu2017, ben2021}. 
The differences in observed outcomes and the estimated counterfactual outcomes in the treated units post-treatment are the treatment effects. SCM variants differ primarily in how the counterfactual outcomes are estimated and in the assumptions required. 
In this paper, we focus on standard SCM \citep{abadie2010}, generalized SCM \citep{xu2017}, and augmented SCM \citep{ben2021} given their widespread use and accessibility through existing software packages.

Standard SCM constructs a synthetic control unit as a convex combination of control units (often called "donors" in this literature), choosing non-negative weights that sum to one to closely match pre-treatment outcomes. The donor pool is the set of control units used to estimate these weights, and the resulting synthetic control is the weighted average of these units. The synthetic control unit's post-treatment outcomes are the estimated counterfactuals. The method is designed for settings with only a single treated unit, where there are many pre-treatment periods and the treated unit lies within the convex hull of the donor pool. However, it can perform poorly when pre-treatment fit is imperfect, i.e., when the treated unit cannot be well approximated by the controls in the donor pool. Uncertainty is typically quantified using placebo (permutation-based) inference, which compares the estimated treated effect for the treated unit to the distribution of effects obtained by reassigning the treatment to control units \citep{abadie2010}.

Standard SCM can also be formulated as a latent factor model \citep{abadie2010}, and many modern approaches explicitly use this latent factor representation for counterfactual estimation. Generalized SCM, which estimates counterfactuals via a low-rank factor model \citep{xu2017}, is one such approach. In particular, the approach models the observed outcomes under control using an interactive fixed effects (IFE) latent factor model, and uses the model to estimate counterfactuals. Compared with standard SCM, this factor modeling approach has two advantages: it accommodates multiple treated units with staggered treatment timing and does not require treated units to lie within the convex hull of the donor units. 

Augmented SCM \citep{ben2021} provides a middle ground by combining standard SCM weighting with a regression-based bias correction. When standard SCM cannot achieve good pre-treatment fit, augmented SCM adjusts for any residual imbalance using an outcome model (typically ridge regression). This improves the pre-treatment alignment of the synthetic controls and the treated units' trends and reduces bias, but introduces additional model dependence and tuning \citep{ben2021}. Augmented SCM also accommodates multiple treated units with staggered treatment timing and relaxes the convex hull condition.

\subsection{Panel data structure and notation}
We consider a balanced panel of units $i = 1,\dots, N$ observed over time periods $t = 1,\dots,T$. 
We assume that each unit-time observation falls into one of three possible policy states: control, resources in place (RIP), or mandate in place (MIP). In the control state, a unit is neither eligible for early adoption nor subject to the policy mandates. In the RIP state, the resources are in place to enable partial or complete early adoption of policy-related measures but compliance is not yet required, e.g., the voluntary access period of PDMPs.
In the MIP state, compliance with the policy is mandated and a unit is assumed to have fully implemented the policy-required measures. 

Let $A_{it}$ be a binary indicator equal to 1 if unit $i$ is in the RIP state at time $t$, indicating that the unit is eligible to early adopt, and $0$ otherwise. Let $D_{it} \in \{0,1\}$ indicate whether unit $i$ is in the MIP state at time $t$. 
We assume that a mandate can only be put in place after or concurrently with the introduction of the necessary resources. Therefore, the RIP state must be initiated prior to or concurrent with the MIP state (i.e., $D_{it} = 1$ implies $A_{it} = 1$). Both the RIP and MIP states are also assumed to be absorbing states, so once a unit enters either state, it cannot revert back during follow-up. Then, a unit with $\left\lbrace A_{it}=0,D_{it}=0 \right\rbrace$ is in the control state, a unit with $\left\lbrace A_{it}=1,D_{it}=0 \right\rbrace$ is in the RIP state, and a unit with $\left\lbrace A_{it}=1,D_{it}=1 \right\rbrace$ is in the MIP state.

We index potential outcomes by time since RIP and time since MIP. Let $T_{i}$ denote the time at which unit $i$ enters the MIP state ($T_i=\infty$ for units that never enter the MIP state during follow-up) and define the event time $E_{it} = \text{max}\left\lbrace 0, t - T_{i}+1\right\rbrace$ as time relative to MIP initiation. Similarly, let $P_{i}$ denote the time at which unit $i$ enters the RIP state ($P_i \leq T_i$, and $P_i=\infty$ for units that never enter the RIP state during follow-up). Then we can define $G_{it} = \text{min}\left\lbrace \text{max}\left\lbrace 0, t - P_i +1 \right\rbrace,T_i-P_i \right\rbrace$ as the time relative to RIP initiation, whose value is truncated at the final time period prior to the mandate being put into place. Control periods have $g=0$ and $e=0$. The RIP period has $\left\lbrace g\in[1,T_i-P_i], e = 0\right\rbrace$ and the MIP period has $\left\lbrace g=T_i-P_i, e \geq 1\right\rbrace$. Let $Y_{it}(g,e)$ represent the potential outcome for unit $i$ at time $t$ when it is $g$ periods from RIP and $e$ periods from MIP. We impose the stable unit treatment value assumption (SUTVA), under which the realized outcome can be written as $Y_{it} = Y_{it}(g,e)$ when $G_{it} = g$ and $E_{it} = e$. Under this assumption, potential outcomes for unit $i$ depend only on its own RIP and MIP path and not those of other units. We then define early adoption effects for $g \geq 1$ and mandate effects for $e \geq 1$ as differences in potential outcomes relative to the control state. 

\subsection{Causal effects of interest}
We focus on the total policy effect and its decomposition into the early adoption effect and the mandate effect. These quantities are defined as functions of the potential outcomes, $Y_{it}(g, e)$. 

First, we define the total policy effect at $e$ periods post-mandate as $$\Delta(e) = \mathbb{E}[Y_{it}(T_i-P_i, e) - Y_{it}(0,0)|G_{it}= T_i - P_i, E_{it} = e], e \geq 1.$$ This estimand compares observed outcomes under both the RIP and MIP states with the untreated counterfactual for treated units. By construction, the total policy effect can be decomposed into early adoption and mandate components. Specifically, $$\Delta(e) = \underbrace{\beta(T_i-P_i)}_{\text{early adoption}} \ + \ \underbrace{\tau(e)}_{\text{mandate}},$$ 
where $\beta$ and $\tau$ are formally defined below. The derivation of this decomposition is provided in Supplementary Material, Section S1. The early adoption effect at $e$ periods post-mandate is evaluated at $g = T_i - P_i$, the realized duration of the RIP period, because $G_{it}$ is truncated at $T_i - P_i$ throughout the MIP period, as described above. Thus, for $e \geq 1, g$ is fixed for each treated unit and is not an additional index of the decomposition. Although $T_i - P_i$ may vary across treated units, the expectations that define $\Delta(e)$ and $\tau(e)$ average over this unit-specific variation, so it is not indexed explicitly. The total policy effect captures the overall impact of the policy intervention on the outcome, including both changes that occur due to the resources being put into place and additional changes that occur due to the mandate. The early adoption and mandate components are defined as follows. We define the early adoption effect at $g$ periods after entry into the RIP state as $$\beta(g) = \mathbb{E}[Y_{it}(g,0)-Y_{it}(0,0)|G_{it} = g, E_{it} = 0], g \geq 1,$$ which represents the expected difference in potential outcomes after $g$ periods of early adoption eligibility relative to the control state, \textit{prior to the mandate}. Next, we define the mandate effect at $e$ periods after entry into the MIP state as $$\tau({e}) = \mathbb{E}[Y_{it}(T_i-P_i,e)-Y_{it}(T_i-P_i,0)|G_{it} = T_i-P_i, E_{it} = e], e \geq 1$$ for units that have a mandate during follow-up. This quantity captures the additional change in the outcome \textit{due to the mandate}, isolating it from any prior changes due to early adoption. 

In many applications, the total policy effect is the primary quantity of interest because it captures the combined impact of the early adoption and mandate effects relative to the counterfactual of no policy action. However, decomposing the total effect into early adoption and mandate components can provide important policy insights. For example, policymakers may be interested in whether the availability of policy-related resources or infrastructure, together with voluntary adoption, is sufficient to achieve the desired effect, or whether a mandate and enforcement are necessary for the policy to be effective. Therefore, the mandate effect may be the most informative when interpreted relative to the early adoption effect, rather than as a standalone quantity.

To achieve identification of these causal effects, we use several assumptions. \\\\
\textbf{Assumption 1 (Model specification)} The outcome generating process can be expressed as a low-rank IFE model, where latent factors capture time-varying unobserved heterogeneity. The number of factors is treated as known. Specifically, the assumed model specification is as follows
\begin{equation}\label{assump1}
Y_{it} = \mu_i + \delta_t + \boldsymbol{\lambda}_i^\text{T} \boldsymbol{f}_t + \boldsymbol{\gamma}^\text{T} X_{it} + \beta(g)A_{it} + \tau(e)D_{it} + \epsilon_{it}.
\end{equation}
The first and second terms, $\mu_i$ and $\delta_t$ are unit and time fixed effects. The $\boldsymbol{f}_t$ is a low-dimensional vector of common factors, and $\boldsymbol{\lambda}_i$ are the corresponding unit-specific factor loadings (both unobserved). $\boldsymbol{\gamma}$ is a vector of regression coefficients corresponding to the vector of observed covariates, $X_{it}$. The function, $\beta(g)$ for $g \geq 1$, represents the average early adoption effect among units in the RIP state, where $A_{it}$ is the corresponding binary indicator of whether unit $i$ is in the RIP state at time $t$ (early adoption eligibility). The average mandate effect among units in the MIP state is denoted by $\tau(e)$ for $e \geq 1$ and $D_{it}$ is an indicator of whether unit $i$ is in the MIP state at time $t$. Lastly, $\epsilon_{it}$ is an idiosyncratic error term, capturing random noise. Moving forward, we let $Z_{it} := (X_{it}, f_t, \lambda_i)$ capture observed covariates $X_{it}$ and latent factors/loadings $f_t$, $\lambda_i$. \\\\
\textbf{Assumption 2 (Positivity)} $P(G_{it} \geq 1 | Z_{it}) \in (0,1) \text{ and } P(E_{it} \geq 1 | Z_{it}) \in (0,1)$ for all $i, t$.  \\\\
\textbf{Assumption 3 (Latent Exchangeability)} $Y_{it}(g,e) \ind (G_{it}, E_{it}) | Z_{it}$. Conditional on the latent factor structure and measured covariates, potential outcomes are independent of the timing of resource availability and mandate implementation. \\\\
\textbf{Assumption 4 (Early adoption and mandate effect structure)} Early adoption effects are a function of time since entry into the RIP state, with a common shape across units. Consistent with the definition of $G_{it}$, the early adoption effect remains fixed at its final pre-mandate value upon transition to the MIP state. In addition, the mandate effect depends only on time since entry into the MIP state and not on the duration of the early adoption (RIP) period.

\subsubsection{Aggregate measures}\label{sec:agg_measures}
In addition to the event time-specific effect measures defined above, it is often convenient to have effect measures aggregated across time. Thus, we define averages of the event time-specific effects, $\beta(g), \tau(e),$ and $\Delta(e)$, over $g$ and $e$ to obtain a single causal measure for each quantity. Let $K_{\text{early}} := \{g: g \geq 1 \}$ denote the times in the RIP period, and let $K_{\text{post}} := \{e: e \geq 1 \}$ denote the times in the MIP period. We define the following aggregate measures:
$$\begin{aligned}
\bar{\Delta}_{\text{post}} &= \frac{1}{|K_{\text{post}}|}\sum_{e \in K_{\text{post}}}\Delta(e) \\
\bar{\beta}_{\text{early}} &= \frac{1}{|K_{\text{early}}|}\sum_{g \in K_{\text{early}}}\beta(g) \\
\bar{\tau}_{\text{post}} &= \frac{1}{|K_{\text{post}}|}\sum_{e \in K_{\text{post}}}\tau(e) 
\end{aligned}$$
Here, $\bar{\Delta}_{\text{post}}$ captures the average total policy effect (during the MIP period), $\bar{\beta}_{\text{early}}$ summarizes average early adoption effects (during the RIP period), and $\bar{\tau}_{\text{post}}$ summarizes average mandate effects (during the MIP period).

\subsection{Early adoption-aware two-stage estimation procedure}\label{sec:estimation_procedure}

We propose a two-stage estimation procedure that separately estimates the early adoption and mandate components of the total policy effect in settings with staggered policy timing. In the first stage, we estimate the early adoption component using pre-mandate data. In the second stage, we residualize outcomes by removing the estimated early adoption component and apply SCM to the residualized outcomes to estimate the mandate component of the total policy effect. We then recover estimates of the total policy effect by combining the estimated early adoption and mandate components. \\\\
\noindent \textbf{Stage 1: Estimating the early adoption component}
\newline \noindent To estimate the early adoption component of the total policy effect, we model pre-mandate outcomes as a function of event time since RIP initiation, $\beta(g)$, using a monotonic spline basis representation interacted with an indicator of whether unit $i$ was in RIP stage at time $t$, $A_{it}$. This interaction term allows us to capture systematic changes in the outcome for units that are eligible for early adoption, using units that have no policy in place during the observed study period as controls. The spline basis provides a flexible representation of the common temporal pattern of early adoption effects across eligible units.
Specifically, we fit the following model using all mandate-free observations ($E_{it} = 0$):
\begin{equation}
    Y_{it} = \mu_i + \delta_t + \boldsymbol{\lambda}_i^{\text{T}}\mathbf{f}_t + \boldsymbol{\gamma}^{\text{T}} X_{it} + \beta(G_{it})A_{it} + \epsilon_{it}
\end{equation}
Here, $\mu_i$, $\delta_t$, $\boldsymbol{\lambda}_i^{\text{T}}\mathbf{f}_t$, and $X_{it}$ are as defined in Equation \ref{assump1}. The number of latent factors may be specified a priori or selected using a data-driven procedure, as described in Section \ref{sec:implement}. The primary quantity of interest is $\beta(g)$, which represents the early adoption effect over event time. We estimate $\beta(g)$ over the full range of observed $g$ values.
\\\\
\noindent \textbf{Stage 2: Estimating the mandate component and total policy effect}
\newline \noindent In the second stage, we remove the estimated early adoption component from the observed outcomes for all units and time periods, and use the resulting residualized outcomes to construct counterfactual outcomes in SCM. Define the residualized outcome as $$\tilde{Y_{it}} = Y_{it} - \hat{\beta}(G_{it})A_{it}.$$ This transformation removes the contribution of early adoption from treated units under the additive effects assumption (Assumption 1), while leaving control units unchanged by construction, ensuring that pre- and post-mandate observations are on a comparable scale when constructing counterfactuals. Next, we apply an SCM variant-- either augmented SCM or generalized SCM (see Section~\ref{sec:implement} below)-- to the residualized outcomes. This yields estimated counterfactual outcomes for each treated unit in the residualized space, denoted $\tilde{Y_{it}}(G_{it}, 0)$. These counterfactuals represent outcomes after removing the early adoption component. We can then map these counterfactuals back to the original outcome scale by reintroducing the early adoption component we previously subtracted out as:
$$\hat{Y}_{it}(G_{it}, 0) = \tilde{Y_{it}}(G_{it}, 0) + \hat{\beta}(G_{it}).$$ 
For the MIP period ($e \geq 1$), our estimator of the mandate effect for treated units at $G_{it} = T_i - P_i$ and $E_{it} = e$ is:
$$\begin{aligned}
\hat{\tau}(e) &= \frac{1}{N_{e}}\sum_{i,t: E_{it} = e} \hat{\tau}(E_{it}) \\
&= \frac{1}{N_{e}}\sum_{i,t: E_{it} = e} Y_{it} - \hat{Y}_{it}(G_{it}, 0) \\
&= \frac{1}{N_{e}}\sum_{i,t: E_{it} = e} Y_{it} - [\tilde{Y_{it}}(G_{it}, 0) + \hat{\beta}(G_{it})] \\
&= \frac{1}{N_{e}}\sum_{i,t: E_{it} = e}[Y_{it} - \hat{\beta}(G_{it})] - \tilde{Y_{it}}(G_{it}, 0) \\
&= \frac{1}{N_{e}}\sum_{i,t: E_{it} = e}\tilde{Y}_{it} - \tilde{Y_{it}}(G_{it}, 0),
\end{aligned}$$
which captures changes in the outcome after the mandate takes effect, net of early adoption. The estimator for the total policy effect among treated units with $e$ treated periods is then given by
$$\begin{aligned}
\hat{\Delta}(e) &= \frac{1}{N_{e}} \sum_{i,t: E_{it} = e} \hat{\Delta}(E_{it}) \\
&= \frac{1}{N_{e}} \sum_{i,t: E_{it} = e} Y_{it} - \hat{Y}_{it}(0,0) \\
&= \frac{1}{N_{e}} \sum_{i,t: E_{it} = e} Y_{it} - \hat{Y}_{it}(G_{it}, 0) + \hat{Y}_{it}(G_{it},0) - \hat{Y}_{it}(0,0) \\
&= \frac{1}{N_{e}} \sum_{i,t: E_{it} = e} \hat{\tau}(E_{it}) + \hat{\beta}(G_{it}) \\
&= \hat{\tau}(e) + \hat{\beta}(T_i-P_i).
\end{aligned}$$ This quantity reflects the combined effect of early adoption and mandate policy components. Then it is straightforward to form the aggregate measures described in \ref{sec:agg_measures} by averaging $\hat{\beta}(g), \hat{\tau}(e)$, and $\hat{\Delta}(e)$ over the values of $g$ and $e$ for which each quantity is well-defined. 

\subsubsection{Pre-treatment fit diagnostics}\label{pre_trt_defn}
While $\tau(e)$ is defined for post-mandate periods, we use placebo estimates to assess model fit during periods in which neither early adoption nor mandate effects should be present, i.e. during control periods ($g = 0$ and $e = 0$). Let $K_{\text{pre}}$ denote the set of pre-mandate event times corresponding to these control periods. We define the average placebo effect as  
$$\bar{\tau}_{\text{pre}} = \frac{1}{|K_{\text{pre}}|}\sum_{e \in K_{\text{pre}}}\hat{\tau}(e).$$ Under adequate model fit, $\bar{\tau}_{\text{pre}}$ should be close to 0. 

\subsubsection{Uncertainty quantification}\label{sec:bootstrap_proc}
We use a non-parametric, unit-level bootstrap to obtain standard errors and 95\% confidence intervals (CI). While the bootstrap procedure is applicable to both the event time-specific and aggregate effect estimates, we focus on inference for the average early adoption effect, the average mandate effect, and the average total policy effect. The bootstrap procedure is as follows: \\
\textbf{Step 1 (Resampling)} For each bootstrap replicate $b = 1, \dots, B$, we sample $n$ units with replacement from the observed set of units and retain the full time series for each sampled unit. This yields a bootstrap panel that preserves within-unit dependence as well as variation in RIP and MIP initiation times. \\
\textbf{Step 2 (Estimation)} On the bootstrap panel, we re-run the two-stage estimation procedure to obtain $\hat\beta^{(b)}(g), \hat\tau^{(b)}(e), \text{ and } \hat\Delta^{(b)}(e)$. For each bootstrap replicate, we compute the corresponding aggregate effect measure by averaging over the relevant event time window. For example, the average mandate effect is computed as $$\hat{\bar\tau}^{(b)}_{\text{post}} = \frac{1}{|K_{\text{post}}|}\sum_{e \in K_{\text{post}}}\hat\tau^{(b)}(e).$$ \\
\textbf{Step 3 (Standard errors and confidence intervals)} After repeating steps 1 and 2 for $b = 1,\dots, B$, we compute the bootstrap standard error for each aggregate effect measure as the sample standard deviation of its bootstrap replicates, e.g. $\{\hat{\bar\tau}^{(1)}_{K_{\text{post}}},\dots, \hat{\bar\tau}^{(B)}_{K_{\text{post}}}\}$ for the average mandate effect. We construct 95\% CIs using Efron's percentile method applied to the bootstrapped distribution. 

\subsection{Implementation}\label{sec:implement}
In the general case with multiple treated units, the second stage of estimation can be done using either generalized SCM or augmented SCM. Below, we implement our approach using each of these methods via the corresponding software packages: \textit{gsynth} for generalized SCM \citep{gsynth} and \textit{augsynth} for augmented SCM \citep{augsynth}. A practical limitation of applying these estimators in settings with staggered adoption is that their event time estimates of the mandate effect, $\tau(e)$, are not directly comparable. In particular, the methods use different indexing schemes and report estimates over different subsets of event times.

The \textit{gsynth} implementation typically covers a symmetric or near-symmetric range around zero based on the earliest MIP date, whereas the \textit{augsynth} implementation defines event time relative to the latest MIP date. For example, when $T = 30$ and the earliest MIP date in the sample is $\text{min}(T_i) = 16$ as in the simulation study, the output for \textit{gsynth} is indexed from $e \in \{-14, \dots, 15 \}$. Here, $e \in [-14, 0]$ denotes pre-mandate event times and $e \in [1,15]$ denotes post-mandate event times. If the latest MIP date is $\text{max}(T_i) = 21$, \textit{augsynth} reports post-mandate event times as $e \in \{0,\dots, 9\}$ and the pre-mandate event times are $e \in \{-20, \dots, -1\}$. As a result, the sets of event times for which $\tau(e)$ is available using generalized SCM and augmented SCM generally differ. Therefore, averages over post-mandate event times are not directly comparable unless the corresponding support ($K_{\text{post}}$) is aligned across methods. 

To enable comparison, we re-index the \textit{augsynth} estimates so that the first treated period corresponds to $e = 1$, matching the indexing used in the \textit{gsynth} implementation. We then identify the set of post-mandate event times where both methods report $\tau(e)$. All aggregate effect measures are computed using only this common set of event times to ensure that the corresponding estimands are comparable when evaluating simulation performance.

In the application, however, the default \textit{augsynth} implementation generates only a single mandate effect estimate because the latest mandate date coincides with the final period of follow-up in the study period. As a result, constructing a meaningful common event time support would require discarding a large portion of the \textit{gsynth} estimates. We therefore present the \textit{gsynth} results, without re-indexing, as our primary estimates and report the corresponding \textit{augsynth} results in Supplementary Material, Section S3. 

In the simulation study, we fix the number of factors at its true value. For the empirical application, we select the spline specifications (polynomial degree and degrees of freedom) and number of latent factors using a data-driven selection procedure. For each combination of candidate values in a finite grid, we fit the IFE model in the first stage of analysis and compute an information criterion. We then select the combination of the number of factors, polynomial degree, and spline degrees of freedom that minimizes this criterion. 
The number of factors is held fixed in the second stage of analysis. 

\section{Simulation study}\label{sec: simulation_study}
We conducted a simulation study to evaluate the performance of the early adoption-aware two-stage estimation procedure in comparison to standard approaches that ignore the presence of early adoption. Below, we describe the data generating process and simulation setup in detail.

\subsection{Data generating process}\label{sec:dgp}
We generate balanced panels with $N = 50$ units observed over $T = 30$ time periods. We assume without loss of generality, units $i = 1,\dots, 20$ have a policy become effective during follow-up ($n_\text{trt} = 20$) while units $i = 21,\dots, 50$ never have a policy become effective ($n_\text{ctr} = 30)$. For each unit with a policy, we draw a time $T_i$ uniformly from $\{16,\dots, 21\}$ that represents the MIP state entry date while units who never enter the MIP state are assigned $T_i = \infty$. As defined previously, time relative to the MIP entry date is $E_{it} = \text{max}\left\lbrace 0, t - T_{i}+1\right\rbrace$ so that $e = 0$ corresponds to the pre-mandate period and $e = 1$ corresponds to the first period in which the mandate is in place. We assume that units enter the RIP state 10 time periods before their MIP entry time, so for each unit $i$ with a policy during follow-up, $P_i = T_i - 10$. Then the time relative to RIP entry is $G_{it} = \text{min}\left\lbrace \text{max}\left\lbrace 0, t - P_i +1 \right\rbrace,10 \right\rbrace$, so that $G_{it} = \{0,1,\dots,10\}$ indexes the control period and the 10 periods in which the unit is eligible for early adoption, respectively. 

According to the model specification given in Equation \ref{assump1}, the untreated potential outcomes are generated from an IFE model: 
\begin{equation*}
    Y_{it}(0,0) = \mu_i + \delta_t + \boldsymbol{\lambda}_i^{\text{T}}\mathbf{f}_t + \epsilon_{it},
\end{equation*}
where $\mu_i \sim N(0,2^2)$ are unit fixed effects and $\delta_t \sim N(0, 0.5^2)$ are time fixed effects. 
In simulations, data are generated using two latent factors, and we do not include any observed covariates, $X_{it}$. Lastly, $\epsilon_{it} \sim N(0,1)$ is an error term. We generate the latent factors, $\boldsymbol{f}_t$, in two ways. In Simulations 1-2, both factors are drawn from independent standard normal distributions. In Simulations 3-6, the first factor follows a smooth time trend plus noise, with the second factor drawn from an independent standard normal distribution. 

The treated potential outcome $g$ periods after RIP entry and $e$ periods after MIP entry is given by
$$\begin{aligned}
Y_{it}(g,e) &= Y_{it}(0,0) + \beta(G_{it})A_{it} + \tau(E_{it})D_{it} \\
&= Y_{it}(0,0) + \beta(G_{it}) + \tau(E_{it}).
\end{aligned}$$
The mandate effect is constant over time following entry into the MIP state: 
$$\tau(e) = 
\begin{cases}
    0, e \leq 0 \\
    20, e \geq 1.
\end{cases}$$
For the early adoption effect, we assume that all units share a common pattern. The baseline early adoption effect function is 
$$\beta^*(g) = 
\begin{cases}
0, g \leq 5 \\
2(g-5), g = 6,\dots, 10.
\end{cases}
$$
We build on this baseline function to define the different $\beta(g)$ functions used in the simulation scenarios described below. The baseline early adoption effect function is zero during the first five eligible periods, reflecting a lower likelihood of early adoption further from the MIP date, and then increases smoothly over $g = 6, \dots, 10$ corresponding to the final five periods before the policy goes into effect. To allow for unit-level heterogeneity in this baseline early adoption pattern, we define $\tilde{\beta}(G_{it}) = s_i\cdot \beta^*(G_{it})$, where $s_i$ are unit-specific scalars generated from $s_i \sim \text{Uniform}(0.9, 1.25)$. 

We consider simulation scenarios both with and without collinearity between the early adoption effects and the latent factor structure. To impose this collinearity in the data generating process, we generate the first factor to follow a smooth time trend (with the remaining factors drawn from an independent standard normal distribution) and then perturb non-zero values of $\beta_{it}$ with the first factor ${f}_{t1}$ and its unit-specific loading ${\lambda}_{i1}$. These scenarios (described in detail below) allow us to evaluate how well the proposed estimation procedure recovers the true early adoption effect when it is correlated with non-policy-related factors driving the outcome.

We consider six simulation scenarios, which we refer to as Simulation 1-Simulation 6 throughout: 
\begin{itemize}
\item \textbf{Simulation 1 (No early adoption)} Early adoption effects are set to zero, so that $\beta(g) = 0$ for all $g$. 
\item \textbf{Simulation 2 (Baseline early adoption with no factor collinearity)} Early adoption follows the pattern $\beta^{\text{ptb}}(G_{it})=\tilde{\beta}(G_{it})$ as defined above, with full early adoption among all eligible units and no additional perturbation based on the latent factors.
\item \textbf{Simulation 3 (Additive time-varying shift)} Early adoption is perturbed by a smooth time trend in the first latent factor: $\beta^{\text{ptb}}(G_{it}) = \tilde{\beta}(G_{it}) + 0.3{f}_{t1}$.
\item \textbf{Simulation 4 (Additive factor-based shift)} Early adoption is perturbed by unit-specific loadings on the first latent factor: $\beta^{\text{ptb}}(G_{it}) = \tilde{\beta}(G_{it}) + 0.3\lambda_{i1}\bar{f}_{1}$, where $\bar{f}_{1}$ is the mean of the first latent factor.
\item \textbf{Simulation 5 (Multiplicative factor-based shift)} Early adoption is scaled by the magnitude of the first latent factor loading: $\beta^{\text{ptb}}(G_{it}) = (1+0.3|\lambda_{i1}|)\tilde{\beta}(G_{it})$.
\item \textbf{Simulation 6 (Factor-based rescaling)} Early adoption is constructed as the interaction between the first latent factor and its unit-specific loading: $\beta^{\text{ptb}}(G_{it}) = 0.5|\lambda_{i1}|{f}_{t1}$.
\end{itemize}
Lastly, the early adoption effect is held constant in the post-mandate time periods at its value in the final pre-treatment (pre-mandate) period ($E_{it} = 0$) by Assumption 4. Thus, for each scenario we define the final $\beta(g)$ as
$$\beta(G_{it})=
\begin{cases}
\beta^{\text{ptb}}(G_{it}), e \leq 0 \\
\beta^{\text{ptb}}(G_{i(T_i - 1)}), e \geq 1. \\
\end{cases}$$
In our main simulation study, all treated units that are eligible for early adoption are assumed to fully early adopt. In a secondary set of simulations, we consider a setting in which only 50\% of treated units early adopt, despite being eligible. We then consider a setting in which, among early-adopting units, 50\% only ``partially'' early adopt ($s_i \sim \text{Uniform}(0.25, 0.75)$) and the remaining 50\% ``fully'' early adopt ($s_i \sim \text{Uniform}(0.9, 1.25)$). These scenarios introduce additional heterogeneity in the degree of early adoption compliance and effectively create multiple ``versions'' of early adoption, resulting in departures from the SUTVA assumptions. Although this violates the strict SUTVA assumption, such heterogeneity is plausible in real-world settings so we assess how our approach performs under such scenarios. Further details and results for these secondary simulations are given in Supplementary Material, Section S2.2. 

\subsection{Estimands}\label{sim_estimands}

As described in Section \ref{sec:agg_measures}, for each effect of interest, we estimate the overall effect as the average of the estimated event time-specific effects over the corresponding event time window. Our primary quantity of interest is the average total policy effect, $\hat{\bar{\Delta}}_{\text{post}}$. We also report the corresponding average early adoption effect, $\hat{\bar{\beta}}_{\text{early}}$, and average mandate effect, 
$\hat{\bar{\tau}}_{\text{post}}$. Recall that for the mandate effect, $\tau(e)$, due to differences in implementation between generalized SCM and augmented SCM, we average only over the common event time support shared by both methods. Finally, we report $\hat{\bar{\tau}}_{\text{pre}}$, the average placebo effect defined in Section \ref{pre_trt_defn} as a diagnostic of pre-treatment fit. 

\subsection{Comparator method and performance metrics}
On each simulated dataset, we apply the early adoption-aware two-stage estimation procedure for each of generalized SCM and augmented SCM, separately. As the corresponding ``conventional'' comparator method, we omit the first stage of our estimation procedure and fit generalized SCM and augmented SCM directly to the observed outcome $Y_{it}$ regressed only on the treatment indicator $D_{it}$, ignoring the possibility of early adoption. From these model fits, we obtain naive analogues of the target causal effects by averaging the differences between treated and counterfactual outcomes over different portions of the study period. Specifically, the naive average early adoption effect is defined as the average difference of treated and counterfactual outcomes over the hypothetical early adoption period, i.e., the ten RIP periods. Because this model does not distinguish early adoption from mandate effects, it does not identify the incremental mandate effect. Instead, the average difference between treated and counterfactual outcomes over all MIP periods is interpreted as the naive average total policy effect since treated units have already accumulated any early adoption effects when the mandate takes effect. The naive average placebo effect is computed as the difference between treated and counterfactual outcomes, averaged across all periods prior to the early adoption (RIP) window.

For each simulation scenario, we generate $R = 200$ replicate datasets from the data generating process described above. For each dataset, we estimate the aggregate effect measures described in Section \ref{sim_estimands} using our early adoption-aware approach and their corresponding naive analogues. Inference is conducted using $B = 500$ bootstrap replicates for each $r$. For each quantity and its counterpart, we compute the bootstrapped standard error and corresponding 95\% CI. Using the known true values of the estimands, we assess performance across replicates in terms of (percent) bias and coverage of 95\% CIs.

\subsection{Results}
\begin{figure}[]
    \includegraphics[width=1\textwidth]{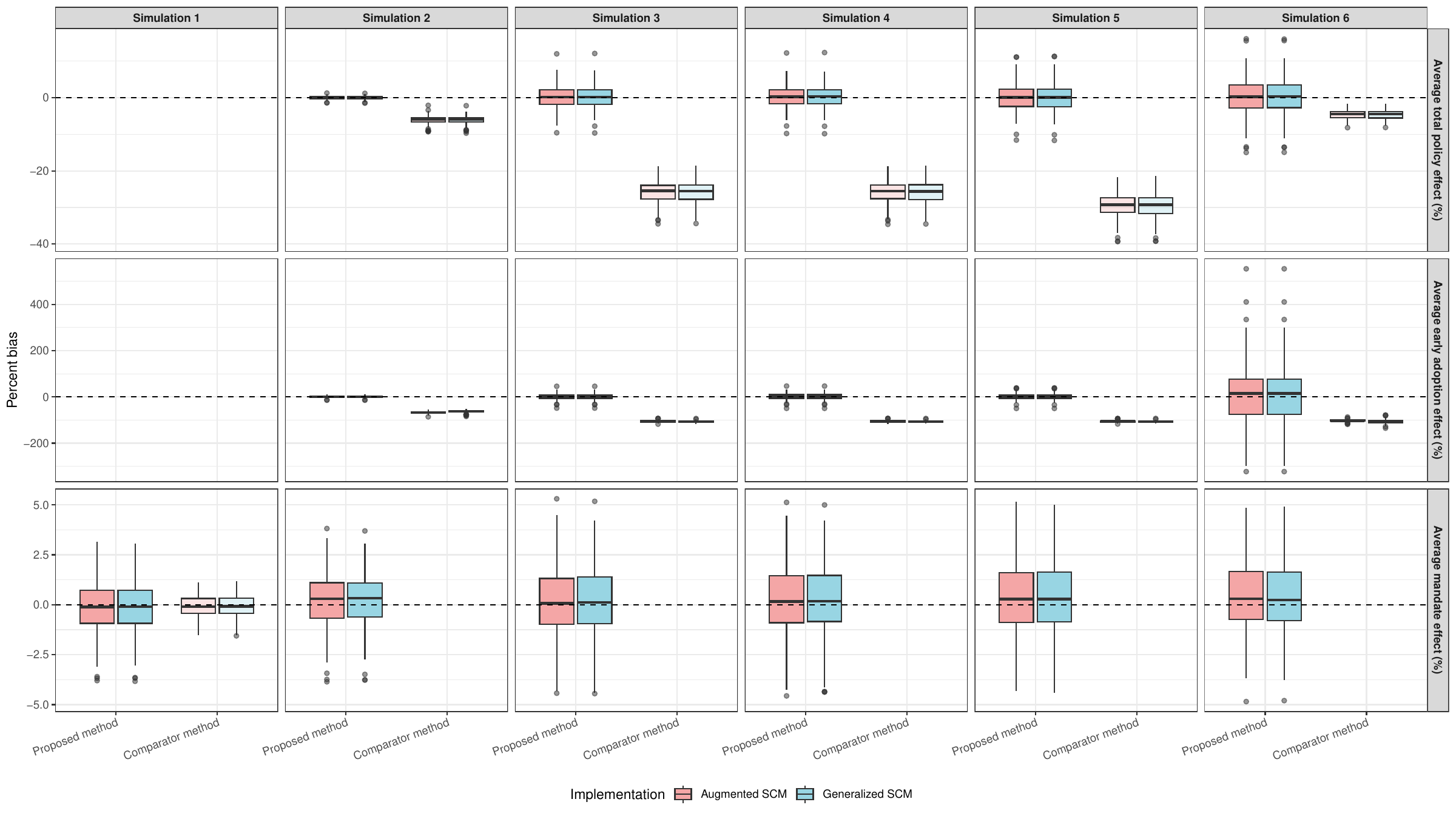}
    \caption{Distribution of percent bias for average total policy, early adoption, and mandate effect estimates in Simulations 1-6. Results are shown for the proposed (two-stage estimator) and comparator (naive estimator) methods, each implemented using augmented SCM and generalized SCM.}
    \label{sim_bias}
\end{figure}

Simulation 1 serves as a benchmark when there is no early adoption effect present. As a result, the total policy effect and mandate effect coincide, so only the average mandate effect is shown in Figure \ref{sim_bias}. In this setting, all methods produce unbiased or near-unbiased estimates of the average mandate effect. For our method, average mandate effect percent bias is approximately 0.15\% and for the standard (non-adjusted) methods, average mandate effect percent bias is approximately -0.08\%. As expected, our method exhibits greater variability in average mandate effect estimates (standard error [SE] = 1.3) relative to standard approaches (SE = 0.56), reflecting the additional noise introduced by residualizing outcomes when no adjustment is necessary. Coverage rates are generally consistent across methods, ranging from 0.93 to 0.97 (Table \ref{coverage}). 

For Simulations 2-6, comparator results are shown only for the average total policy effect and average early adoption effect because the naive estimator does not identify the mandate effect. Simulation 2 introduces a smooth time trend into the factor structure and serves as the baseline setting with early adoption present, but without additional collinearity between early adoption effects and latent factors. In this setting, the two-stage estimator produced nearly unbiased estimates for all average effects, with boxplots centered close to zero (Figure \ref{sim_bias}). In contrast, the naive estimator for augmented SCM and generalized SCM displayed negative bias in estimates of the average total policy effect (-6.0\% for both) and especially for the average early adoption effect (-67.7\% and -62.3\%). 
These biases are reflected in the coverage probabilities: for average early adoption, average mandate effect, and average total policy effect, our methods achieve coverage close or equal to 0.95, whereas coverage for the comparator method is near zero (0.0) as shown in Table \ref{coverage}. 

Simulations 3-6 apply the four factor-based perturbations described in \ref{sec:dgp}, inducing varying degrees of collinearity between early adoption effects and the latent factor structure. Across the first three perturbations (Simulations 3-5), our method accurately recovers the average total policy effect, the average early adoption effect, and the average mandate effect (Figure \ref{sim_bias}). The comparator method reveals similar patterns to those in Simulation 2 (Figure \ref{sim_bias}) with increased magnitudes of bias for the average total policy effect in particular. 
Differences between the two naive estimators are minimal across all estimands. Biases for the average total policy effect are nearly identical under the naive estimators. For the average early adoption effect, the augmented SCM naive estimator yields marginally smaller bias than the generalized SCM naive estimator. Our method achieves near-nominal coverage for the average early adoption effect, average mandate effect, and average total policy effect (0.95-0.98) while coverage for the comparator method is close to zero (0.0-0.05) across Simulations 3-5 (Table \ref{coverage}). 

In Simulation 6, early adoption effects are rescaled based on the first latent factor. In this setting, our method exhibits greater variability than the comparator method for the average early adoption effect (SE: 137 vs 4.5 for augmented SCM; 137 vs 8.55 for generalized SCM). The increase in variability was smaller for the average total policy effect (SE: 6.3 vs 1.7). Our method continues to yield near-unbiased estimates for the average total policy effect, average early adoption effect, and average mandate effect. The comparator method still fails to recover the average early adoption effect, showing substantial negative bias (mean percent bias of -103\% for augmented SCM naive estimator and -107\% for generalized SCM naive estimator). However, the naive estimators perform substantially better for the total policy effect than in earlier simulations. 
The naive estimator has a mean percent bias of -4.6 for the average total policy effect, while the two-stage estimator remains nearly unbiased. This improvement is expected because the early adoption effect is relatively small in this scenario, so ignoring early adoption introduces minimal bias into the estimated total policy effects. In this setting, the proposed method exhibits over-coverage for the average total policy effect (0.98-0.99), whereas coverage for the comparator method remains close to zero (Table \ref{coverage}). Bias in the estimated average placebo effect for each simulation scenario and estimator is reported in Supplementary Materials, Section S2.1.
         
\begin{table}[htbp]
\centering
\begin{tabular}{ccccc}
\hline
Simulation & Method & $\bar{\Delta}_{\text{post}}$ & $\bar{\beta}_{\text{early}}$ & $\bar{\tau}_{\text{post}}$  \\
\hline
1 & Two-stage estimator (Augmented SCM)    &    NA    & NA        &  \bf{0.94}    \\
1 & Two-stage estimator (Generalized SCM)   &    NA  & NA &   0.93   \\
1 & Naive estimator (Augmented SCM) &    NA  & NA       &     \bf{0.96}    \\
1 & Naive estimator (Generalized SCM)  &    NA    & NA       &   0.97      \\
\hline
2 & Two-stage estimator (Augmented SCM)   & \bf{1.0}        & \bf{0.94} & 0.94 \\
2 & Two-stage estimator (Generalized SCM)  & \bf{1.0}  & \bf{0.94}       & \bf{0.95} \\
2 & Naive estimator (Augmented SCM) & 0.0        & 0.0       & NA      \\
2 & Naive estimator (Generalized SCM) & 0.0        & 0.0       & NA       \\
\hline
3 & Two-stage estimator (Augmented SCM) & \bf{0.97}  & \bf{0.97}    & \bf{0.95} \\
3 & Two-stage estimator (Generalized SCM)    & 0.98  & 0.98 &  \bf{0.95} \\
3 & Naive estimator (Augmented SCM) & 0.0        & 0.0   & NA      \\
3 & Naive estimator (Generalized SCM) & 0.0        & 0.0    & NA     \\
\hline
4 & Two-stage estimator (Augmented SCM)    & \bf{0.97}  & \bf{0.97}       & \bf{0.95} \\
4 & Two-stage estimator (Generalized SCM)    & \bf{0.98}  & 0.98 & \bf{0.95}       \\
4 & Naive estimator (Augmented SCM) & 0.0        & 0.0       & NA      \\
4 & Naive estimator (Generalized SCM)& 0.0        & 0.0       & NA    \\
\hline
5 & Two-stage estimator (Augmented SCM)    & \bf{0.98}  & \bf{0.97} & \bf{0.96}       \\
5 & Two-stage estimator (Generalized SCM)    & \bf{0.98}  & 0.98 & \bf{0.96} \\
5 & Naive estimator (Augmented SCM) & 0.0        & 0.0       & NA     \\
5 & Naive estimator (Generalized SCM) & 0.0        & 0.0       & NA      \\
\hline
6 & Two-stage estimator (Augmented SCM)    & \bf{0.98}  & \bf{0.99} & \bf{0.99}       \\
6 & Two-stage estimator (Generalized SCM)    & 0.99  & \bf{0.99}       & 1.0 \\
6 & Naive estimator (Augmented SCM) & 0.06        & 0.0 & NA     \\
6 & Naive estimator (Generalized SCM) & 0.06        & 0.0 & NA      \\
\hline
\end{tabular}
\caption{Simulated coverage probability for our method using generalized SCM and augmented SCM, compared with their naive analogues. Results are reported by simulation setting and average effect (total policy, early adoption, and mandate). Within each simulation and effect type, the value(s) closest to 0.95 are highlighted in bold.}
\label{coverage}
\end{table}

\section{Application}
Figure \ref{policy_plot} shows that 49 states and the District of Columbia had operational PDMP access gradually, beginning in the early 2000s. The earliest access occurred in Kentucky in the third quarter of 2001, with most states gaining access by the early-to-mid 2010s. Missouri is the only state that did not have an operational PDMP during the study period. We also exclude Delaware because the policy dataset indicates that the mandated-use requirement was enacted before operational PDMP access, an ordering that is inconsistent with our method's assumption that the RIP state precedes (or is concurrent with) the MIP state. Mandated use of PDMPs generally followed several years after they first became available, creating long RIP periods in many states with earlier access such as Kentucky, Michigan, Maine, New Mexico, and Utah. Among states that acted more recently, the RIP periods are generally shorter, and some states had mandated use in place within a few years of gaining access. Overall, there is substantial staggered timing across states and several years in which early adoption might exist through accessible, but not yet required, PDMPs. 

\begin{figure}
    \centering
    \includegraphics[width=\linewidth]{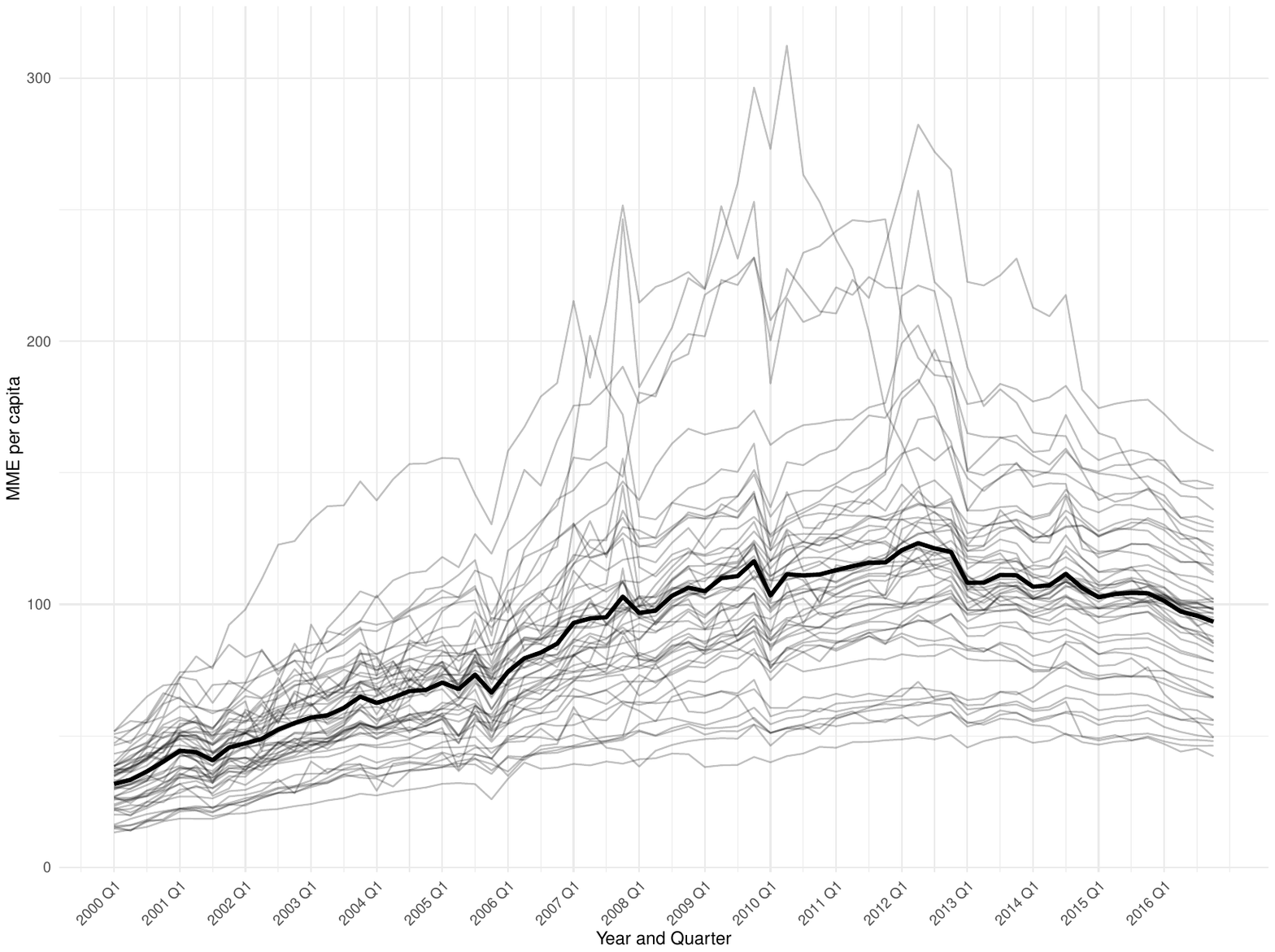}
    \caption{State-level opioid dispensing trends, measured in MME per capita, from 2000-2016 by quarter. Gray lines represent individual states while the solid black line denotes the national average.}
    \label{opioid_trends}
\end{figure}

State-level trends in quantity of hydrocodone and oxycodone distributed, measured in MME per capita, from 2000 to 2016 are shown in Figure \ref{opioid_trends}. Opioid dispensing increases gradually across nearly all states during the early 2000s, with more rapid increases for some states in the mid 2000s. Dispensing levels peak between 2010 and 2012, followed by a steady decline through the end of the study period. While the overall trend is generally similar across states, there is substantial heterogeneity in the magnitude of per capita dispensing, particularly around the peak in 2010.  

\subsection{Results}\label{app_results}
\begin{figure}[h]
    \centering
    \includegraphics[width=0.90\linewidth]{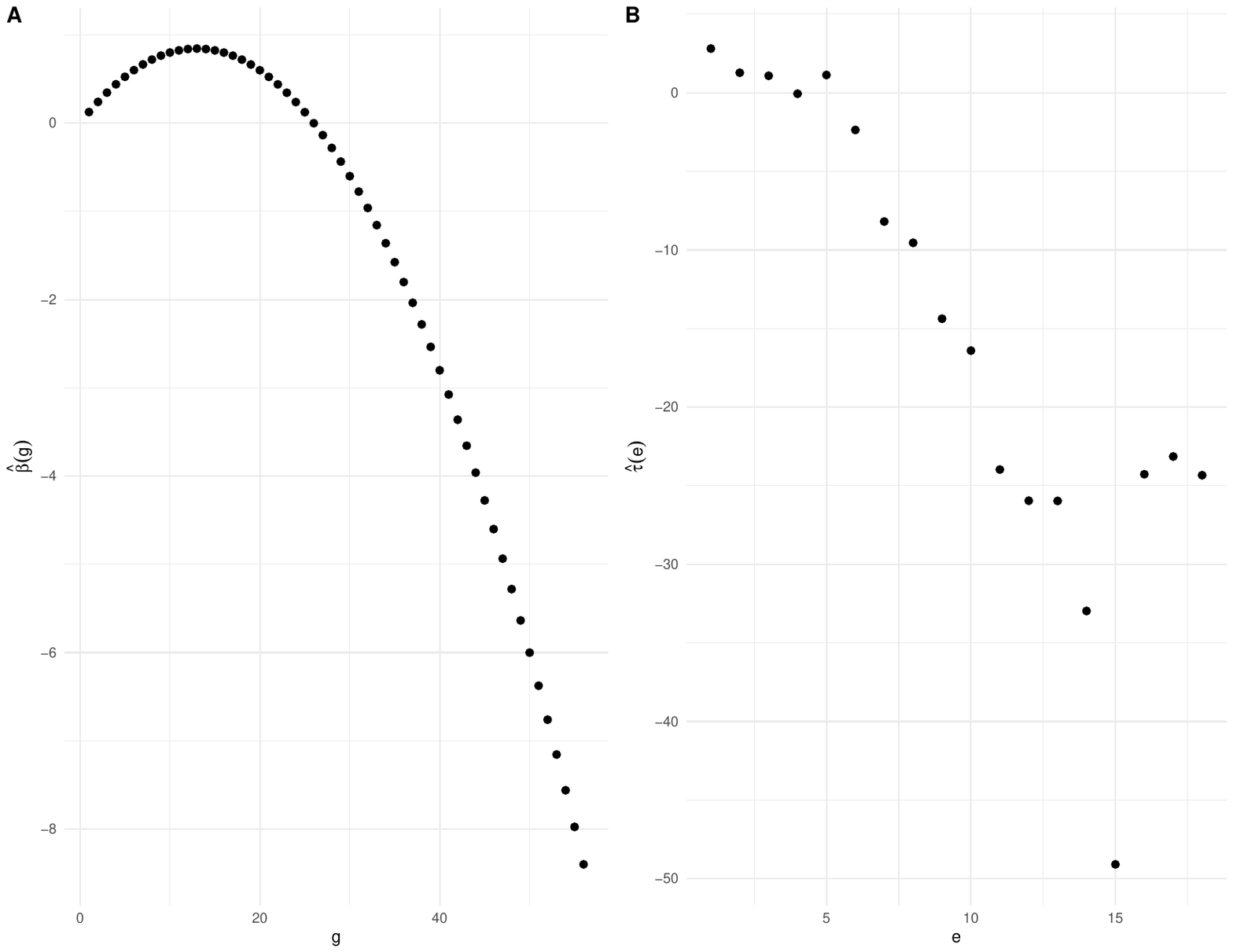}
    \caption{Panel A shows the estimated early adoption effects, $\hat{\beta}(g)$, of state-level PDMP policies on MME per capita opioid dispensing from hydrocodone and oxycodone as a function of $g$ (time since RIP state entry). Panel B shows the corresponding estimated mandate effects, $\hat{\tau}(e)$, as a function of $e$ (time since MIP state entry).}
    \label{beta_tau_primary}
\end{figure}

The results of our analysis of state-level PDMP policies and opioid dispensing using our proposed approach implemented via generalized SCM are presented below. The estimated average placebo effect was small in magnitude (0.03; 95\% CI: -0.09, 0.49), indicating that the model adequately explains pre-treatment trends in the outcome. The estimated average total policy effect, combining early adoption and mandate components, was $-23.6$ (95\% CI: -38.0 to 2.8). The point estimate corresponds to an average reduction of 23.6 MME of opioids dispensed per capita per quarter relative to the estimated counterfactual in the absence of the policy. The estimated average early adoption effect over the RIP period was negative but with large uncertainty (-1.7; 95\% CI: -10.1, 11.9). Although the 95\% CI includes zero, the negative point estimate is consistent with modest reductions in opioid dispensing during the period when PDMPs were accessible but prescriber queries were not yet mandated. Panel A of Figure \ref{beta_tau_primary} shows the early adoption effect estimate for each RIP period event time. We observed a gradual decline in $\hat{\beta}(g)$ as a function of $g$ (time since entry into the RIP state), with the final pre-mandate estimate, $\hat{\beta}(56)$ equal to -8.4. Panel B of Figure \ref{beta_tau_primary} shows the estimated mandate effect at each MIP event time. The estimates generally indicate a steady decline in MME per capita following mandate implementation, reaching a plateau in later periods. The estimated average mandate effect was $-15.2$ (95\% CI: -38.7, 8.3). While the estimated average mandate effect was larger in magnitude than the early adoption effect, it was also not statistically significant. Note that the average early adoption and average mandate effects do not sum to the average total policy effect because each aggregate estimand is defined as an average over a different event time window. Overall, while the point estimates suggest reductions in MME per capita due to PDMP policy implementations, the estimated effects lack statistical significance.

Corresponding results using augmented SCM were similar in magnitude and precision (see Supplementary Material, Section S3 for more details). Sensitivity analyses using alternative specifications for spline polynomial degree, degrees of freedom, and the number of latent factors are provided in Supplementary Material, Section S4. Our results were robust to these alternative specifications, with no substantive changes observed. 

\section{Discussion}
In this paper, we aimed to quantify the impact of state-level PDMP laws on opioid dispensing in the United States. This task is complicated by the fact that PDMP laws usually allow for a period of access and voluntary use, i.e., early adoption of the PDMP system, prior to mandating use. Existing quasi-experimental methods for estimating policy effects from panel data typically assume that pre-policy outcomes are unaffected by the policy, which is violated when early adoption is present. Motivated by this, we developed a novel two-stage early adoption-aware estimation procedure that decomposes the total policy effect into early adoption and mandate components within a potential outcomes framework for multiple units with staggered policy timing. Our two-stage procedure first fits an IFE model with a spline in time since policy-related resources become available to estimate a common early adoption function and then residualizes outcomes with this estimated component before applying generalized or augmented SCM to estimate mandate and total policy effects. By explicitly modeling and subtracting early adoption effects before constructing synthetic controls, our framework provides a principled way to separate ``anticipatory'' behavior from mandate-related changes and to obtain more accurate, interpretable estimates of total policy impact. 

In our application to state-level PDMP policies and pharmacy-based opioid dispensing, point estimates indicated modest reductions in MME of opioids dispensed per capita during the period in which PDMPs were accessible, but not required for prescribers (average early adoption effect) and larger negative average total policy effects after combining early adoption and mandate components. However, all corresponding confidence intervals were wide and included the null. These findings are consistent with several plausible explanations.

First, although the negative point estimates during the voluntary access period are consistent with education efforts surrounding PDMP rollout (e.g., onboarding and targeted training as systems become available), several barriers may still inhibit use of the system. Multiple studies cite time burden, log-in and access difficulties, and poor electronic health record (EHR) integration as key obstacles to routine PDMP use \citep{alpert2024, martin2021barriers}. EHR integration studies further show that when one-click access is introduced, the number of queries increases significantly \citep{weiner2021opioid, neprash2022effect}, but these increases in querying behavior may only translate into modest or inconsistent changes in opioid dispensing at the population level.

Second, assessing only overall opioid doses dispensed (per capita) could conceal clinically meaningful changes in prescribing patterns. MME per capita may decrease because fewer patients receive opioids, because prescribers shift toward lower-MME substances, or because use of higher-MME substances decline, but it can also remain flat if changes offset each other. For example, if the PDMP leads clinicians to stop prescribing to some lower-dose or short-term patients while maintaining or increasing doses among high-need patients, the total number of opioid prescriptions could decrease while the average MME per prescription increases enough such that aggregate MME per capita is unchanged. Likewise, substitution away from higher-MME substances towards other opioids can mask clinically relevant shifts when all dispensing is aggregated into a single metric \citep{balestra2021, wen2019, dowell2022cdc}. 

Third, there is substantial heterogeneity in mandate scope, enforcement, and technological features in state-level PDMP policies. States differ in which providers are covered, the clinical situations that trigger a required query, which drugs and refills are included, and whether requirements are enforced with active auditing and penalties or complaints-based systems \citep{sun2024, stone2020implementation}. They also vary in key implementation features such as real-time reporting, data sharing, EHR integration, and delegate access \citep{martin2021barriers}. Multi-state studies consistently find stronger reductions under broad, comprehensive mandates with robust enforcement and better technology than under narrower or weakly enforced policies \citep{wen2019}. Together, these mechanisms suggest that the negative, but imprecise estimates we obtain are consistent with the complex ways in which PDMP policies may affect opioid dispensing, including heterogeneity across states and providers as well as changes in prescribing patterns that may not be fully captured by quarterly MME per capita. 

While our work contributes a novel estimation procedure for using SCM in the presence of early adoption, it has limitations. First, we assume that early adoption follows a common shape in time over the RIP period across treated units and that the mandate effect depends only on time since entry into the MIP state. Together, these assumptions imply that the total policy effect can be expressed as an additive decomposition with no interaction between time since MIP initiation, $e$, and the duration of the RIP period, $T_i - P_i$. We additionally assume that the outcome generating process is correctly specified as a low-rank IFE model. Violations of these modeling assumptions, including substantially different early adoption patterns across treated states or mandate effects that vary with the duration of the RIP period, could bias estimates. In settings where early adoption patterns are highly heterogeneous, it may be more appropriate to apply our method separately to individual treated states or subgroups of homogeneous units. Residual confounding that is not captured by the latent factor structure could also bias estimates. Our approach also relies on the SUTVA assumption. Although we assess the robustness of our method to violations that arise from heterogeneous versions of early adoption in a secondary simulation study, we do not consider interference or spillover effects, such as cross-state changes in opioid prescribing or dispensing induced by neighboring states' PDMP policies. Second, our uncertainty quantification relies on a unit-level bootstrap, whose finite-sample properties may be sensitive to the number of treated units and staggered policy timing. Third, in our application to PDMPs, policy dates are given at the month-year level while opioid dispensing data was aggregated to the quarterly level, which can obscure smaller, incremental changes. 



\begin{acks}[Acknowledgments]
The computations in this paper were run on the FASRC Cannon cluster supported by the Faculty of Arts and Sciences Division of Science Research Computing Group.
\end{acks}

\begin{funding}
This work was supported by National Institutes of Health grant T32ES007142.
\end{funding}

\section*{Conflicts of interest}
The authors have no conflicts of interest to disclose.

\section*{Data and code availability}
State-level PDMP policy data are provided in the supplementary materials of \citep{lee2021}. ARCOS data are publicly available \citep{DEA_ARCOS}. Code for the analysis and visualizations is available at \url{https://github.com/sarika1999/early-adoption-scm}.

\begin{supplement}
\stitle{Appendices. }
\sdescription{Additional appendices are provided containing (1) the derivation of the effect decomposition, (2) additional simulation results including violations of the SUTVA assumption, (3) additional application results, and (4) sensitivity analyses for the application.}
\end{supplement}

\bibliographystyle{imsart-nameyear}
\bibliography{bib}

\end{document}


\begin{frontmatter}
\title{Supplementary materials for ``Evaluating the effects of policy interventions subject to early adoption: A case study of prescription drug monitoring programs and opioid dispensing''}
\runtitle{}
\begin{aug}
\author[A]{\fnms{Sarika}~\snm{Aggarwal}\ead[label=e1]{saggarwal@g.harvard.edu}}
\author[A]{\fnms{Brent A.}~\snm{Coull}\ead[label=e2]{bcoull@hsph.harvard.edu}}
\author[A]{\fnms{Nima}~\snm{Hejazi}\ead[label=e3]{nhejazi@hsph.harvard.edu}}
\author[A]{\fnms{Rachel C.}~\snm{Nethery}\ead[label=e4]{rnethery@hsph.harvard.edu}}
\address[A]{Department of Biostatistics, Harvard T.H. Chan School of Public Health\printead[presep={,\ }]{e1}\printead[presep={,\ }]{e2}\printead[presep={,\ }]{e3}\printead[presep={,\ }]{e4}}
\end{aug}
\end{frontmatter}

\renewcommand{\thesection}{S\arabic{section}}
\renewcommand{\thefigure}{S\arabic{figure}}
\renewcommand{\thetable}{S\arabic{table}}

\section{Decomposition derivation}
\label{decomp}
For units in the MIP state, $G_{it}$ is fixed at $T_i - P_i$, the final event time in the RIP state immediately before the mandate takes effect. Therefore, 
$$\Delta(e) = \mathbb{E}[Y_{it}(T_i - P_i, e) - Y_{it}(0,0) | G_{it} = T_i - P_i, E_{it} = e].$$ Adding and subtracting $Y_{it}(T_i - P_i, 0)$ gives

$$\begin{aligned}
\Delta(e) = &\mathbb{E}[Y_{it}(T_i - P_i, e) - Y_{it}(T_i-P_i,0) | G_{it} = T_i - P_i, E_{it} = e] + \\ 
&\mathbb{E}[Y_{it}(T_i - P_i, 0) - Y_{it}(0,0) | G_{it} = T_i - P_i, E_{it} = e].
\end{aligned}$$
The first term is exactly $\tau(e)$. Under Assumption 4, the early adoption effect remains fixed at its final RIP value upon entry into the MIP state. Then the potential outcome corresponding to $T_i - P_i$ periods of early adoption, in the absence of the mandate, is the same after entry into the MIP state. Mathematically, 
\begin{align*}
&\mathbb{E}\left[Y_{it}(T_i-P_i,0)-Y_{it}(0,0)\right.\\
&\qquad\left.\mid G_{it}=T_i-P_i, E_{it}=e\right]\\
&=\mathbb{E}\left[Y_{it}(T_i-P_i,0)-Y_{it}(0,0)\right.\\
&\qquad\left.\mid G_{it}=T_i-P_i, E_{it}=0\right]\\
&=\beta(T_i-P_i).
\end{align*}
The final equality follows from the definition of $\beta(g)$, evaluated at its final value, $g = T_i - P_i$. Therefore,
$$\Delta(e) =  \ \underbrace{\tau(e)}_{\text{mandate}} + \underbrace{\beta(T_i-P_i)}_{\text{early adoption}} \ .$$

\section{Additional simulation results}

\subsection{Placebo effects}\label{supp:placebo}

\begin{figure}[]
    \includegraphics[width=1\textwidth]{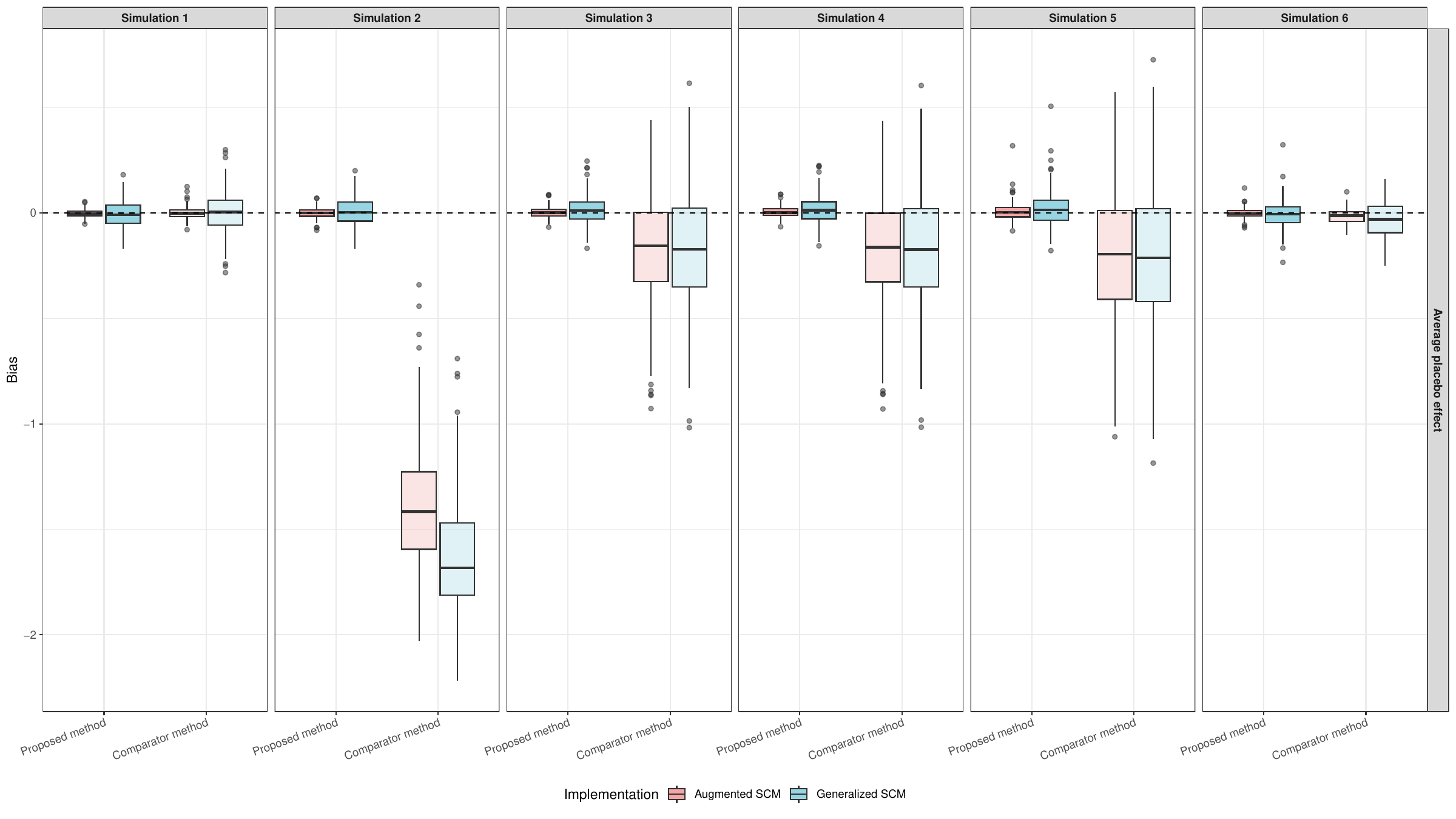}
    \caption{Distribution of bias for average placebo effect estimates in Simulations 1-6. Results are shown for the proposed and comparator method, each implemented using augmented SCM and generalized SCM.}
    \label{sim_bias_supp}
\end{figure}

Figure \ref{sim_bias_supp} shows the bias for the estimated average placebo effect for each scenario in our primary simulation study to assess pre-treatment fit. Across all simulation settings, the estimated average placebo effect for the two-stage estimator was centered near zero, indicating good pre-treatment fit. The naive estimator had little bias in Simulation 1, with poorer pre-treatment fit in Simulations 2, especially, and Simulations 3-5. 

\subsection{SUTVA violations}\label{supp:SUTVA}

\begin{figure}[]
    \includegraphics[width=1\linewidth]{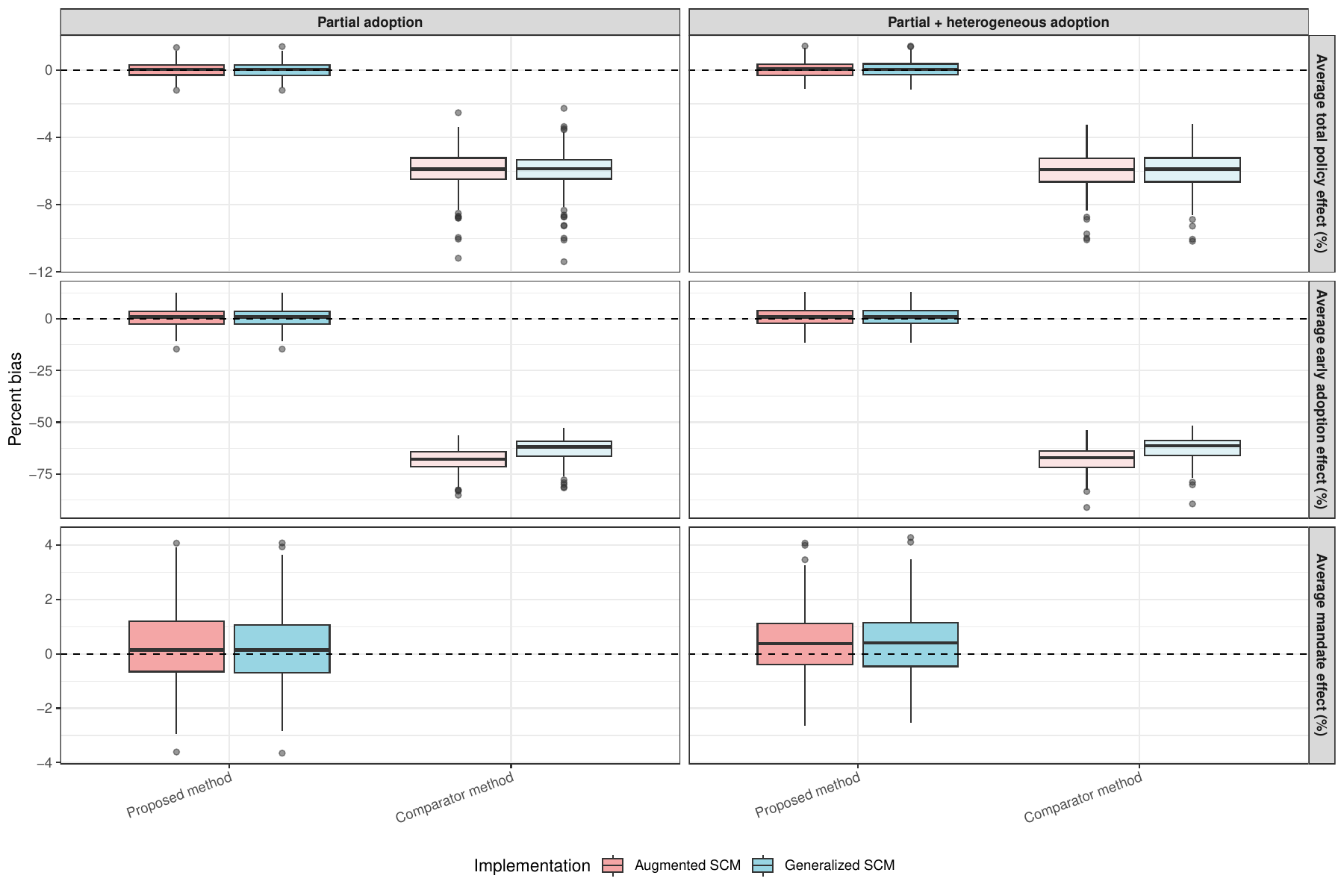}
    \caption{Distribution of percent bias for average total policy, early adoption, and mandate effect estimates in Simulation 2 when only 50\% of eligible units early adopt (first column) and when, among eligible units, 25\% partially early adopt and 25\% fully early adopt (second column). Results are shown for the proposed and comparator method, each implemented using augmented SCM and generalized SCM.}
    \label{sutva_bias_revised}
\end{figure}

In our secondary set of simulations, we build on the scenario described in Simulation 2 (baseline early adoption with no factor perturbation), introducing additional variation in both the number of early adopting units and the degree of early adoption. Figure \ref{sutva_bias_revised} presents the bias for each average effect when only 10 of the 20 eligible units early adopt (partial adoption) and when, among these 10 early adopters, 5 partially adopt and the remaining 5 fully adopt (partial + heterogeneous adoption). Overall performance is similar to that of Simulation 2, where estimates were approximately unbiased across all average effects. Under partial and heterogeneous adoption, mean percent bias increases relative to the partial adoption scenario by approximately 1.5x for the average early adoption and mandate effects and 3x for the average total policy effect. Performance is largely equivalent between the augmented SCM and generalized SCM implementations of the two-stage estimator, and both outperform their corresponding naive estimator. Our proposed method achieves near-nominal coverage (with slight over-coverage), ranging from 0.94 to 1.0 under partial adoption and from 0.96 to 1.0 under partial and heterogeneous adoption (Table \ref{coverage_SUTVA}). Both naive estimators exhibit near-zero coverage in both settings (Table \ref{coverage_SUTVA}).

\begin{table}[htbp]
\centering
\begin{tabular}{ccccc}
\hline
Violation & Method & 
$\bar{\Delta}_{\text{post}}$ &
$\bar{\beta}_{\text{early}}$ & $\bar{\tau}_{\text{post}}$  \\
\hline
Partial adoption & Two-stage estimator (Augmented SCM)    &    \bf{1.0}    & \bf{0.97}        &   0.94     \\
Partial adoption & Two-stage estimator (Generalized SCM)  & \bf{1.0}  & \bf{0.97}  &     \bf{0.95}   \\
Partial adoption & Naive estimator (Augmented SCM) &    0.0  & 0.0     &     NA   \\
Partial adoption & Naive estimator (Generalized SCM) &    0.0    & 0.0        &   NA     \\
\hline
Partial + heterogeneous adoption & Two-stage estimator (Augmented SCM)  & \bf{1.0}  & 0.97       & \bf{0.97} \\
Partial + heterogeneous adoption & Two-stage estimator (Generalized SCM)  & \bf{1.0}  & \bf{0.96}       & \bf{0.97} \\
Partial + heterogeneous adoption & Naive estimator (Augmented SCM) & 0.0        & 0.0       & NA       \\
Partial + heterogeneous adoption & Naive estimator (Generalized SCM) & 0.0        & 0.0      & NA       \\
\hline
\end{tabular}
\caption{Simulated coverage probability for our method using generalized SCM and augmented SCM, compared with their naive analogues under different violations of the SUTVA assumption. Results are reported by simulation setting, violation type, and average effect (early adoption, mandate effect, and total policy effect). Within each simulation and effect type, the value(s) closest to 0.95 are highlighted in bold.}
\label{coverage_SUTVA}
\end{table}

\section{Additional application results}\label{supp:application}
Below we report the results of our primary analysis implemented via augmented SCM, corresponding to the main results in Section 5.1 of the main manuscript. Augmented SCM yields a small average placebo effect (-0.04; 95\% CI: -0.3, 0.06), consistent with good pre-treatment fit. The average total policy effect remains negative but is attenuated towards the null compared to the primary results (-5.0; 95\% CI: -13.5, 11.4). By design, the estimated average early adoption effect is the same as in the main results with a marginally different CI due to the bootstrap procedure (-1.7; 95\% CI: -8.6, 14.5). While the estimated average mandate effect is positive, the 95\% CI is wide and includes 0 (-35.4, 16.6). Overall, these results largely align with those in the main text, suggesting modest reductions in MME per capita but with high uncertainty. 


\section{Sensitivity analyses}\label{supp:sensitivity}

\begin{figure}
    \centering
    \includegraphics[width=0.75\linewidth]{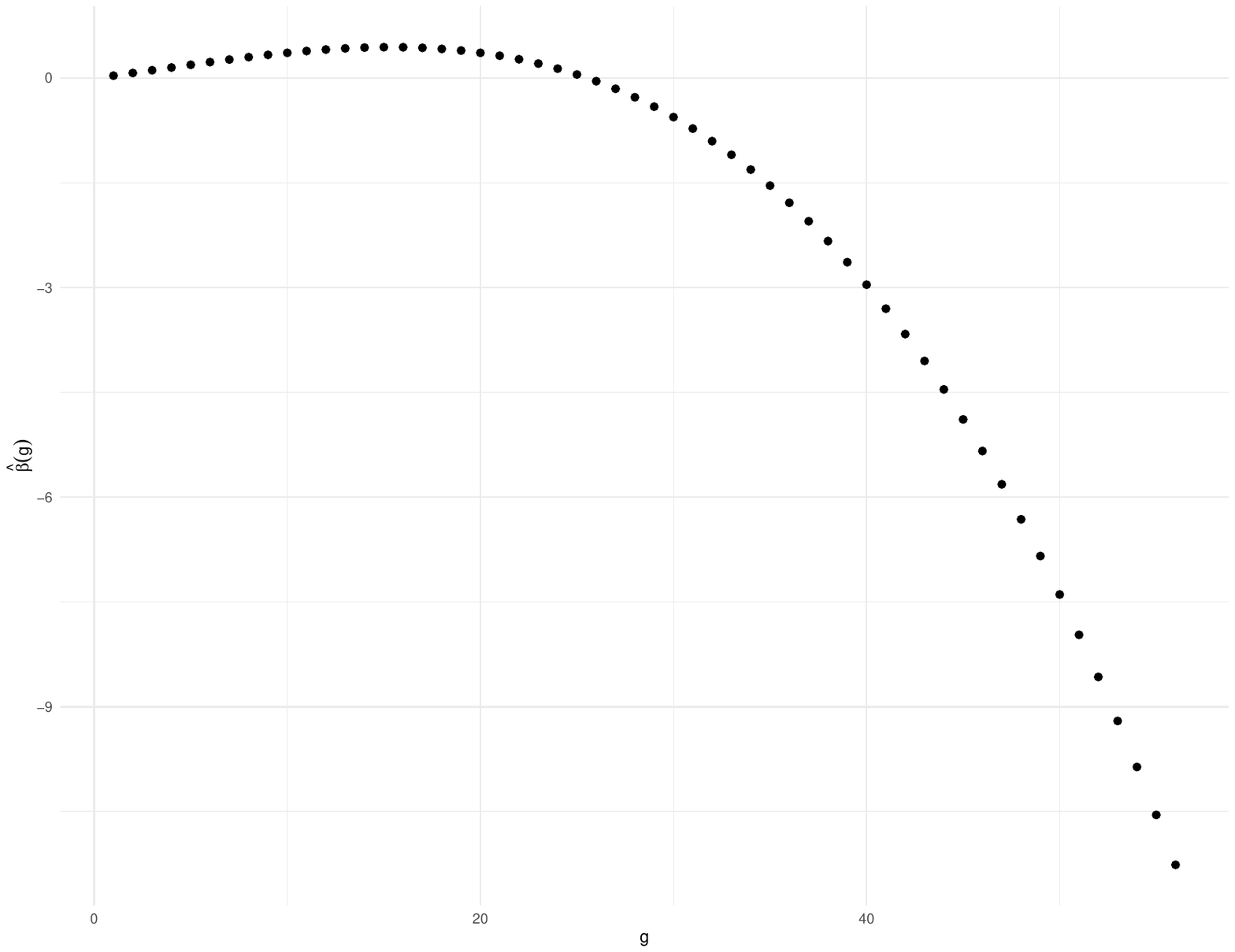}
    \caption{Estimated early adoption effects, $\hat{\beta}(g)$, of state-level PDMP policies as a function of $g$ (time since RIP state entry) for MME per capita opioid dispensing from hydrocodone and oxycodone when $K = 3$, df = 2, and $r = 5$.}
    \label{sens1}
\end{figure}

\begin{figure}
    \centering
    \includegraphics[width=0.75\linewidth]{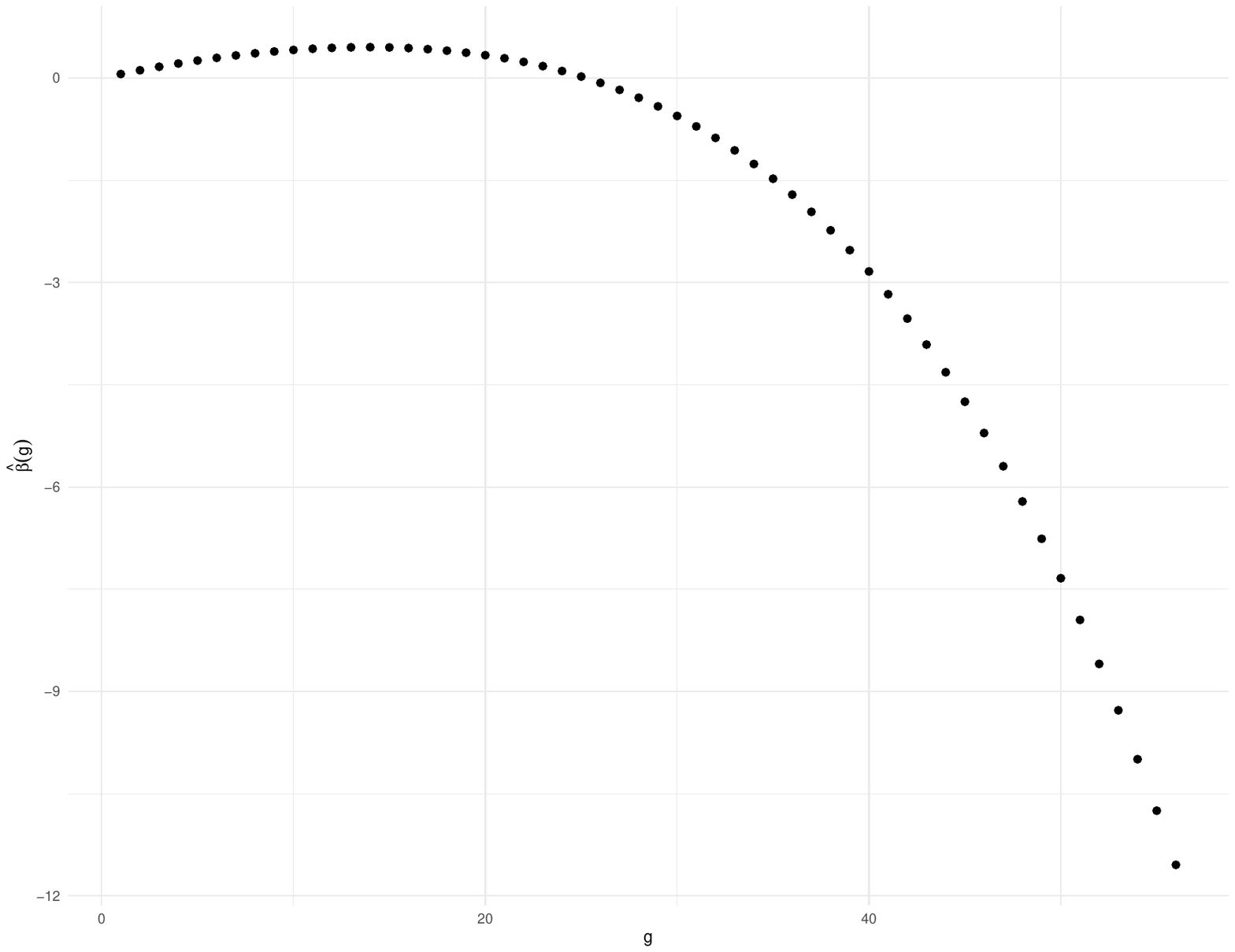}
    \caption{Estimated early adoption effects, $\hat{\beta}(g)$, of state-level PDMP policies as a function of $g$ (time since RIP state entry) for MME per capita opioid dispensing from hydrocodone and oxycodone when $K = 4$, df = 3, and $r = 5$.}
    \label{sens2}
\end{figure}

\begin{figure}
    \centering
    \includegraphics[width=0.75\linewidth]{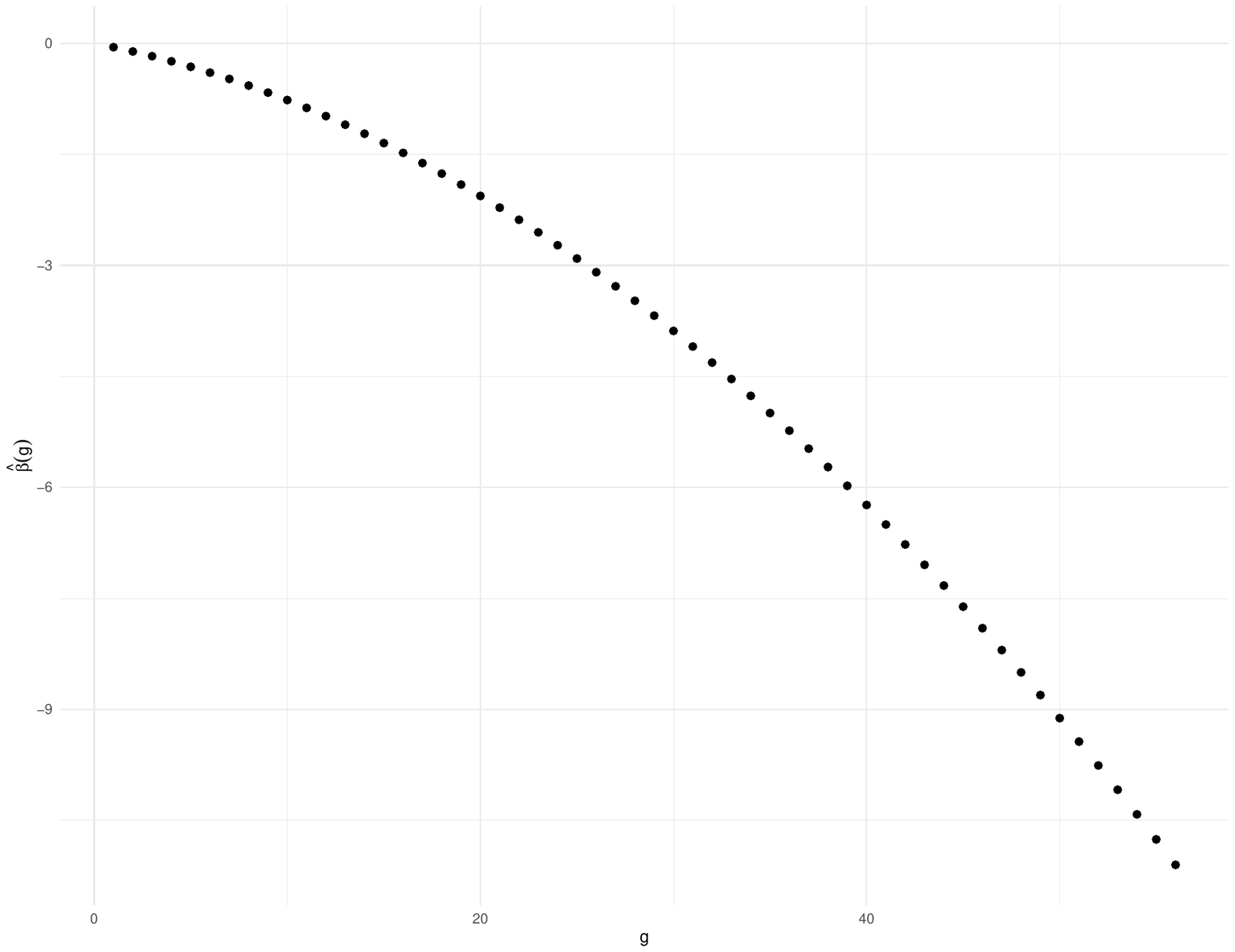}
    \caption{Estimated early adoption effects, $\hat{\beta}(g)$, of state-level PDMP policies as a function of $g$ (time since RIP state entry) for MME per capita opioid dispensing from hydrocodone and oxycodone when $K = 2$, df = 1, and $r = 4$.}
    \label{sens3}
\end{figure}

Table \ref{sens_results} reports the effect estimates and 95\% CIs for MME per capita from dispensed hydrocodone and oxycodone under different values for polynomial degree ($K$), spline degrees of freedom (df), and number of latent factors ($r$). Specific event time estimates for the early adoption effect are shown in Figures \ref{sens1}-\ref{sens3}. Across all specifications, the estimated effects were similar to the main results though with larger standard errors, indicating that our findings are robust to the spline specification and number of latent factors.

\begin{table}[]
    \centering
    \begin{tabular}{|c|c|c|c|c|c|}
    \hline
    $K$ & df & $r$ & $\bar{\Delta}_{\text{post}}$ & $\bar{\beta}_{\text{early}}$ & $\bar{\tau}_{\text{post}}$ \\
    \hline
    3 & 2 & 5 & -24.5 (-39.4, 3.6) & -2.2 (-11.3, 10.8) & -13.2 (-36.8, 10.9) \\
    \hline
    4 & 3 & 5 & -24.5 (-37.7, 2.7) & -2.1 (-9.8, 11.7) & -13.0 (-44.0, 8.1) \\
    \hline
    2 & 1 & 4 & -28.2 (-42.8, 3.5) & -4.3 (-15.4, 13.0) & -17.1 (-47.7, 11.8) \\
    \hline
    \end{tabular} 
    \caption{Estimated average total policy effect, early adoption effect, and mandate effect of state-level PDMP policies on MME per capita opioid dispensing (from hydrocodone and oxycodone) under alternative model specifications using generalized SCM. Corresponding 95\% confidence intervals are reported for each estimand.}
    \label{sens_results}
\end{table}    

